# EFFECT OF PROMOTER ARCHITECTURE ON THE CELL-TO-CELL VARIABILITY IN GENE EXPRESSION

Alvaro Sanchez[1], Hernan G. Garcia[2], Daniel Jones[3], Rob Phillips[3,4], Jané Kondev[5,*]

[1]Graduate Program in Biophysics and Structural Biology and [5]Department of Physics, Brandeis University, Waltham, MA, USA.

[2]Department of Physics, [3]Department of Applied Physics and [4]Department of Bioengineering, California Institute of Technology, Pasadena, CA, USA.

* Corresponding author (kondev@brandeis.edu)

**RUNNING TITLE**: Promoter architecture and cell-to-cell variability

## ABSTRACT

According to recent experimental evidence, the architecture of a promoter, defined as the number, strength and regulatory role of the operators that control the promoter, plays a major role in determining the level of cell-to-cell variability in gene expression. These quantitative experiments call for a corresponding modeling effort that addresses the question of how changes in promoter architecture affect noise in gene expression in a systematic rather than case-by-case fashion. In this article, we make such a systematic investigation, based on a simple microscopic model of gene regulation that incorporates stochastic effects. In particular, we show how operator strength and operator multiplicity affect this variability. We examine different modes of transcription factor binding to complex promoters (cooperative, independent, simultaneous) and how each of these affects the level of variability in transcription product from cell-to-cell. We propose that direct comparison between *in vivo* single-cell experiments and theoretical predictions for the moments of the probability distribution of mRNA number per cell can discriminate between different kinetic models of gene regulation.

## AUTHOR SUMMARY

Stochastic chemical kinetics provides a framework for modeling gene regulation at the single-cell level. Using this framework, we systematically investigate the effect of promoter architecture, that is, the number, quality and position of transcription factor binding sites, on cell-to-cell variability in transcription levels. We compare architectures resulting in transcriptional activation with those resulting in transcriptional repression. We start from simple activation and repression motifs with a single operator sequence, and explore the parameter regime for which the cell-to-cell variability is maximal. Using the same formalism, we then turn to more complicated architectures with more than one operator. We examine the effect of independent and cooperative binding, as well as the role of DNA mechanics for those architectures where DNA looping is relevant. We examine the interplay between operator strength and operator number, and we make specific predictions for single-cell mRNA-counting experiments with well characterized promoters. This theoretical approach makes it possible to find the statistical response of a population of cells to perturbations in the architecture of the promoter, it can be used to quantitatively test physical models of gene regulation *in vivo*, and as the basis of a more systematic approach to designing new promoter architectures.

## INTRODUCTION

A fundamental property of all living organisms is their ability to gather information about their environment and adjust their internal, physiological state in response to environmental conditions. This property, shared by all organisms, includes the ability of single-cells to respond to changes in their environment by regulating their patterns of gene expression. By regulating the genes they express, cells are able to survive changes of the extracellular pH or osmotic pressure, switch the mode of sugar utilization when the sugar content in their medium changes, or respond to shortages in key metabolites by adapting their metabolic pathways. Perhaps more interesting is the organization of patterns of gene expression in space and time resulting in the differentiation of cells into different types, which is one of the defining features of multicellular organisms. Much of this regulation occurs at the level of transcription initiation, and is mediated by simple interactions between transcription factor proteins and DNA, leading to genes being turned on or off. Understanding how genes are turned on or off (as well as the more nuanced expression patterns in which the level of expression takes intermediate levels) at a mechanistic level has been one of the great challenges of molecular biology and has attracted intense attention over the past 50 years.

The current view of transcription and transcriptional regulation has been strongly influenced by recent experiments with single-cell and single-molecule resolution [1,23,4,5,6,7,8,9,10,11,12,13,14,15,16]. These experiments have confirmed the long-suspected idea that gene expression is stochastic [17,18], meaning that different steps on the path from gene to protein occur at random, leading to fluctuations of the gene product in time. This stochasticity also causes variability in the number of messenger RNAs (mRNA) and proteins produced from cell-to-cell in a colony of isogenic cells [3,4,16,19,20]. An important consequence of stochastic gene expression is that the concentrations of proteins and mRNAs inside the cell are fluctuating (*noisy*) quantities, even in steady-state conditions. The question of how transcriptional regulatory networks function reliably in spite of the noisy character of the inputs and outputs has attracted much experimental and theoretical interest [21,22]. A different, but also very relevant, question is whether cells actually exploit this stochasticity to fulfill any physiologically important task. This issue has been investigated in many different cell types and it has been found that stochastic gene expression is a key player in processes as diverse as cell fate determination in the retina of *Drosophila melanogaster* [23], entrance to the competent state of *B. subtilis* [12], resistance of yeast colonies to antibiotic challenge [3], maintenance of HIV latency [24], enabling self-destructive cooperative behavior in bacterial colonies [25] or the induction of the lactose operon in *E. coli* [5]. Other examples have been found, and reviewed elsewhere [26,27]. The overall conclusion of all of these studies is that noise in gene expression can have important physiological consequences in natural and synthetic systems and that the overall architecture of the gene regulatory network can greatly affect the level of stochasticity.

A number of theoretical and experimental studies have revealed multiple ways in which the architecture of the gene regulatory network affects cell-to-cell variability in gene expression. Examples of mechanisms for the control of stochasticity have been proposed and tested, including the regulation of translational efficiency [13], the presence of negative feedback loops [28,29,30], or the propagation of fluctuations from upstream regulatory components [31].

Another important source of stochasticity in gene expression is fluctuations in promoter activity, caused by stochastic association and dissociation of transcription factors, chromatin remodeling events, and formation of stable pre-initiation complexes [4,5,10,20,32]. In particular, it has been reported that perturbations to the architecture of yeast and bacterial promoters, such as varying the strength of transcription factor binding sites[3], the number and location of such binding sites [6,16], the presence of auxiliary operators that mediate DNA looping [5], or the competition of activators and repressors for binding to the same stretch of DNA associated with the  promoter [1], may strongly affect the level of variability.

Our goal is to examine all of these different promoter architectures from a unifying perspective provided by stochastic models of transcription leading to mRNA production. The logic here is the same as in earlier work where we examined a host of different promoter  architectures using thermodynamic models of transcriptional regulation [33,34]. We now generalize those systematic efforts to examine the same architectures, but now from the point of view of stochastic models. Stochastic models allow us to assess the unique signature provided by a particular regulatory architecture in terms of the cell-to-cell variability it produces.

First, we investigate in general theoretical terms how the architecture of a promoter affects the level of cell-to-cell variability. The architecture of a promoter is defined by the collection of transcription factor binding sites (also known as operators), their number, position within the promoter, their strength, as well as what kind of transcription factors bind them (repressors, activators or both), and how those transcription factors bind to the operators (independently, cooperatively, simultaneously). We apply the master-equation model of stochastic gene expression [35,36,37,38]  to increasingly complex promoter architectures [32], and compute the moments of the mRNA and protein distributions expected for these promoters. Our results provide quantitative predictions for the level of variability in gene expression for natural promoters, and can aid in the rational design of synthetic promoters.

The second point of this paper is to make use of stochastic kinetic models of gene regulation to propose *in vivo* tests of the molecular mechanisms of gene regulation by transcription factors that have been proposed as a result of *in vitro* biochemical experiments. The idea of using spontaneous fluctuations in gene expression to infer properties of gene regulatory circuits is an area of growing interest, given its non-invasive nature and its potential to reveal regulatory mechanisms *in vivo*. Different theoretical methods have recently been proposed, which could be employed to distinguish between different modes (AND/OR) of combinatorial gene regulation, and to rule out candidate regulatory circuits [29,39,40] based solely on properties of noise in gene expression, such as the autocorrelation function of the fluctuations [29] or the three-point steady state correlations between multiple inputs and outputs [39,40].

Here, we make experimentally testable predictions about the level of cell-to-cell variability in gene expression expected for different bacterial promoters, based on the physical kinetic models of gene regulation that are believed to describe these promoters *in vivo*. In particular, we focus on how varying the different parameters (i.e. mutating operators to make them stronger or weaker, varying the intracellular concentration of transcription factors, etc.) should affect the level of variability in gene expression. This way, cell-to-cell variability in gene expression is used as a tool for testing kinetic models of transcription factor mediated regulation of gene expression *in vivo*.

The remainder of the paper is organized as follows: First we describe the theoretical formalism we use to determine analytic expressions for the moments of the probability distribution for both mRNA and protein abundances per cell. Next, we examine how the architecture of the promoter affects cell-to-cell variability in gene expression. We focus on simple and cooperative repression, simple and cooperative activation, and transcriptional regulation by distal operators mediated by DNA looping. We investigate how noise in gene expression caused by promoter activation differs from repression, how operator multiplicity affects noise in gene expression, the effect of cooperative binding of transcription factors, as well as DNA looping. For each one of these architectures we present a prediction of cell-to-cell variability in gene expression for a number of bacterial promoters that have been well characterized experimentally in terms of their mean values. These predictions suggest a new round of experiments to test the current mechanistic models of gene regulation at these promoters.

## MATHEMATICAL METHODS

In order to investigate how promoter architecture affects cell-to-cell variability in gene expression, we use a model based on classical chemical kinetics (illustrated in Figure 1A), in which a promoter containing multiple operators may exist in as many biochemical states as allowed by the combinatorial binding of transcription factors to their operators. The promoter transitions stochastically between the different states as transcription factors bind and fall off. Synthesis of mRNA is assumed to occur stochastically at a constant rate that is different for each promoter state. Further, transcripts are assumed to be degraded at a constant rate per molecule.

This kind of model is the kinetic counterpart of the so-called "thermodynamic models" of transcriptional regulation [41], and it is the standard framework for interpreting the kinetics of gene regulation in biochemical experiments, both *in vivo* [5,7] and *in vitro* [42,43]. This class of kinetic models can easily accommodate stochastic effects, and allows the formulation of a master equation from which the properties of the probability distribution of mRNA and protein copy number per cell can be computed. They are often referred to as the standard model of stochastic gene expression [38,44,45]. The degree of cell-to-cell variability in gene expression may be quantified by the stationary variance, defined as the ratio of the standard deviation and the mean of the probability distribution of mRNA or protein copy number per cell [35], or else by the Fano factor, the ratio between the variance and the mean. These two are the two most common metrics of noise in gene expression, and the relation between them will be discussed later.

In order to compute the noise strength from this class of models, we follow the same approach as in a previous article [32], which extends a master equation derived elsewhere [36,37,46] to promoters with arbitrary combinatorial complexity. The complexity refers to the existence of a number of discrete promoter states corresponding to different arrangements of transcription factors on the promoter DNA. Promoter dynamics is described by trajectories involving stochastic transitions between promoter states which are induced by the binding and unbinding of transcription factors. A detailed derivation of the equations which describe promoter dynamics can be found in the supplementary materials, but the essentials are described below.

There are only two stochastic variables in the model: the number of mRNA transcripts per cell ($m$), and the state of the promoter, which is defined by the pattern of transcription factors bound to their operator sites. The promoter state is described by a discrete and finite stochastic variable ($s$) (for an example, see Figure 1A). The example in Figure 1A illustrates the simplest model of transcriptional activation by a transcription factor. When the activator is not bound (state 1), mRNA is synthesized at rate $r_1$. When the activator is bound to the promoter (state 2), mRNA is synthesized at the higher rate rate $r_2$. The promoter switches stochastically from state 1 to state 2 with rate $k_A^{on}$, and from state 2 to state 1 with rate $k_A^{off}$. Each mRNA molecule is degraded with rate $\gamma$.

The time evolution for the joint probability of having the promoter in states 1 or 2, with $m$ mRNAs in the cell (which we write as $p(1,m)$ and $p(2,m)$, respectively), is given by a master equation, which we can build by listing all possible reactions that lead to a change in cellular state, either by changing $m$ or by changing $s$ (figure 1b). The master equation takes the form:

$$\frac{d}{dt}p(1,m) = -k_A^{on} p(1,m) + k_A^{off} p(2,m) - r_1 p(1,m) - \gamma m p(1,m) + r_1 p(1,m-1) + \gamma(m+1)p(1,m+1),$$

$$\frac{d}{dt}p(2,m) = k_A^{on} p(1,m) - k_A^{off} p(2,m) - r_2 p(2,m) - \gamma m p(2,m) + r_2 p(2,m-1) + \gamma(m+1)p(2,m+1) \ .$$

Inspecting this system of equations, we notice that by defining the vector:

$$\vec{p}(m) = \begin{pmatrix} p(1,m) \\ p(2,m) \end{pmatrix}, \tag{1}$$

and the matrices

$$\hat{K} = \begin{bmatrix} -k_A^{on} & k_A^{off} \\ k_A^{on} & -k_A^{off} \end{bmatrix} \ ; \ \hat{R} = \begin{bmatrix} r_1 & 0 \\ 0 & r_2 \end{bmatrix} \ ; \ \hat{I} = \begin{bmatrix} 1 & 0 \\ 0 & 1 \end{bmatrix}, \tag{2}$$

we can rewrite the system of equations (1) in matrix form.

$$\frac{d}{dt}\vec{p}(m) = \left[ \hat{K} - \hat{R} - m\gamma \hat{I} \right] \vec{p}(m) + \hat{R}\,\vec{p}(m-1) + (m+1)\gamma \hat{I}\,\vec{p}(m+1) \ . \tag{4}$$

This has several advantages, but the most important one is that the same matrix equation applies to all of the promoter architectures we consider, so if we can find a solution for the moments of the probability distribution in this case, that solution will be general for any of the other promoter architectures. As we will show below, the matrix approach reduces the task of obtaining analytical expressions for the moments of the steady state mRNA distribution for an arbitrarily complex promoter to solving two simple linear matrix equations (more details are given in the supplementary materials).

The matrices appearing in equation (4) all have simple and intuitive interpretations. The matrix $\hat{K}$ describes the stochastic transitions between promoter states: The off-diagonal elements of the matrix $\{\hat{K}_{ij}\}$ are the rates of making transitions from promoter state $(j)$ to promoter state $(i)$. The diagonal elements of the matrix $\{\hat{K}_{jj}\}$ are negative, and they represent the net probability flux out of state $(j)$ : $\{\hat{K}_{jj}\} = -\sum_{i \neq j} \hat{K}_{ij}$. The matrix $\hat{R}$ is a diagonal matrix whose element $\{\hat{R}_{jj}\}$ gives the rate of transcription initiation when the promoter is in state $(j)$. Finally, the matrix $\hat{I}$ is the identity matrix.

An example of matrices $\hat{K}$ and $\hat{R}$ is presented pictorially in Figures 1C and 1D. It is straightforward to see that even though equation (4) has been derived for a two-state promoter, it also applies to any other promoter architecture. What will change for different architectures are the dimensions of the matrices and vectors (these are given by the number of promoter states) as well as the values of the rate constants that make up the matrix elements of the various matrices.

An important limit of the master equation, which is often attained experimentally is the steady state limit, in which the probability distribution for mRNA number per cell does not change with time. Although the time dependence of the moments of the mRNA distribution can be easily computed from our model, for the sake of simplicity and because most experimental studies have been performed on cells in steady state, we focus on this limit. As shown in the supplementary materials, analytic expressions for the first two moments of the steady state mRNA probability distribution are found by multiplying both sides of equation (4) by $m$ and $m^2$ respectively, and then summing $m$ from 0 to $\infty$. After some algebra (elaborated in an earlier paper and in the SI), we find that the first two moments can be written as:

$$\langle m \rangle = \frac{\vec{r} \cdot \vec{m}_{(0)}}{\gamma}, \qquad (5)$$

$$\langle m^2 \rangle = \langle m \rangle + \frac{\vec{r} \cdot \vec{m}_{(1)}}{\gamma}. \qquad (6)$$

The vector $\vec{r}$ contains the ordered list of rates of transcription initiation for each promoter state. For the two-state promoter shown in Figure 1, $\vec{r} = (r_1, r_2)$. The vector $\vec{m}_{(0)}$ contains the steady state probabilities for finding the promoter in each one of the possible promoter states, while $\vec{m}_{(1)}$ is the steady-state mean mRNA number in each promoter state. The vector $\vec{m}_{(0)}$ is the solution to the matrix equation

$$\hat{K}\,\vec{m}_{(0)} = 0, \qquad (7)$$

while the vector $\vec{m}_{(1)}$ is obtained from

$$\left(\hat{K} - \gamma\,\hat{I}\right)\vec{m}_{(1)} + \hat{R}\,\vec{m}_{(0)} = 0. \qquad (8)$$

Figure 1 illustrates the following algorithm for computing the intrinsic variability of mRNA number for promoters of arbitrarily complex architecture:

1) Make a list of all possible promoter states and their kinetic transitions (Figure 1B)
2) Construct the matrices $\hat{K}$ and $\hat{R}$, and the vector $\vec{r}$, (Figures 1C, D, and E).
3) Solve equations (7-8) to obtain $\vec{m}_{(0)}$ and $\vec{m}_{(1)}$
4) Plug solutions of (7-8) into equations (5-6) to obtain the moments.

The normalized variance of the mRNA distribution in steady state is then computed from the equation:

$$\eta^2 = \frac{Var(m)}{\langle m \rangle^2} = \frac{\langle m^2 \rangle - \langle m \rangle^2}{\langle m \rangle^2} = \frac{1}{\langle m \rangle} + \frac{1}{\langle m \rangle^2} \left( \frac{\vec{r} \cdot \vec{m}_{(1)}}{\gamma} - \langle m \rangle^2 \right) \quad . \tag{9}$$

Equation (9) reveals that, regardless of the specific details characterizing promoter architecture, the intrinsic noise is always the sum of two components, and it can be written as

$$\eta^2 = \frac{1}{\langle m \rangle} + \eta^2_{promoter} \quad . \tag{10}$$

The first component is due to spontaneous stochastic production and degradation of single mRNA molecules, it is always equal to $1/\langle m \rangle$, and is independent of the architecture of the promoter. For an unregulated promoter that is always active and does not switch between multiple states (or does so very fast compared to the rates of transcription and mRNA degradation), the mRNA distribution is well described by a Poisson distribution [45,47], and the normalized variance is equal to $1/\langle m \rangle$. The second component ("promoter noise") results from promoter state fluctuations, and captures the effect of the promoter's architecture on the cell-to-cell variability in mRNA:

$$\eta^2_{promoter} = \frac{1}{\langle m \rangle^2} \left( \frac{\vec{r} \cdot \vec{m}_{(1)}}{\gamma} - \langle m \rangle^2 \right). \tag{11}$$

In order to quantify the effect of the promoter architecture in the level of cell-to-cell variability in mRNA expression, we define the deviation in the normalized variance caused by gene regulation relative to the baseline Poisson noise (see Figure 2):

$$Fold-change\ mRNA\ noise = \frac{\eta^2}{\eta^2_{Poisson}} = \frac{\eta^2_{promoter} + \eta^2_{Poisson}}{\eta^2_{Poisson}} = \frac{Var(m)/\langle m \rangle^2}{1/\langle m \rangle} = \frac{Var(m)}{\langle m \rangle} \tag{12}$$

Therefore, the deviation in the normalized variance caused by gene regulation is equal to the ratio between the variance and the mean. This parameter is also known as the Fano factor. Thus, for any given promoter architecture, the Fano factor quantitatively characterizes how large the mRNA noise is relative to that of a Poisson distribution of the same mean (i.e. how much the noise for the regulated promoter elevates with respect to the Poisson noise). This is the parameter that we will use throughout the paper as the metric of cell-to-cell variability in gene expression.

For proteins, the picture is only slightly more complicated. As shown in the supplementary materials, in the limit where the lifetime of mRNA is much shorter than that of the protein it encodes for (a limit that is often fulfilled [32]), the noise strength of the probability distribution of proteins per cell takes the following form (where we define by $n$ the copy number of proteins per cell):

$$\frac{Var(n)}{\langle n \rangle^2} = \frac{\langle n^2 \rangle - \langle n \rangle^2}{\langle n \rangle^2} = \frac{1+b}{\langle n \rangle} + \frac{1}{\langle n \rangle^2}\left(b\frac{\vec{r}\,\vec{n}_{(1)}}{\delta} - \langle n \rangle^2\right), \tag{13}$$

where $\delta$ stands for the protein degradation rate, and the constant $b$ is equal to the protein burst size (the average number of proteins produced by one mRNA molecule). The mean protein per cell is given by $\langle n \rangle = b\frac{\vec{r}\,\vec{m}_{(0)}}{\delta}$, and the vector $\vec{n}_{(1)}$ is the solution of the algebraic equation:

$$\left(\hat{K} - \delta\,\hat{I}\right)\vec{n}_{(1)} + b\hat{R}\,\vec{m}_{(0)} = 0, \tag{14}$$

The reader is referred to the supplementary materials for a detailed derivation and interpretation of these equations. In the previous section we have shown that the noise for proteins and mRNA take very similar analytical forms. Indeed, if we define $\vec{r}_n = b\,\vec{r}$ and $\hat{R}_n = b\,\hat{R}$, as the vector and matrix containing the average rates of protein synthesis for each promoter state, it is straightforward to see that equations (8) and (14) are really the same equation, with the only difference being that in equation (14) the matrix $\hat{R}_n$ represents the rates of protein synthesis, so all the rates of transcription are multiplied by the translation burst size $b$. Therefore, the vectors $\vec{m}_{(1)}$ and $\vec{n}_{(1)}$ are only going to differ in the prefactor $b$ multiplying all the different transcription rates. We conclude that the promoter contribution to the noise takes the exact same analytical form both for proteins and for mRNA, with the only other quantitative difference being the different rates of degradation for proteins and mRNA. Therefore, promoter architecture has the same qualitative effect on cell-to-cell variability in mRNA and protein numbers. All the conclusions about the effect of promoter architecture on cell-to-cell variability in mRNA expression are also valid for proteins. For the sake of simplicity, and since we do not lose generality, we focus on mRNA noise and distributions.

*Intrinsic and Extrinsic noise*

Although all sources of noise in gene expression contribute to the total cell-to-cell variability, the intrinsic component depends only on the stochastic nature of the chemical reactions that lead to mRNA production and decay, including promoter switching, transcription and degradation, all of which we can account for in the kind of models considered here. On the other hand, the extrinsic component is affected by global cell fluctuations in all of the kinetic rates in the model, which may have complex origins and are difficult to connect to mechanistic models. The cell-to-cell variability due to intrinsic noise sources can be measured directly using recently developed experimental methods [8,20,48]. Recent experimental studies in mammalian cells, yeast and bacteria, have found that most of the cell-to-cell variability in mRNA could be attributed to intrinsic noise sources, such as stochastic promoter activation and repression, and stochastic mRNA synthesis and degradation [2,6,9]. The noise strength computed in this paper is the intrinsic contribution.

*Parameters and Assumptions*

In order to evaluate the equations in our model, we use parameters that are consistent with experimental measurements of rates and equilibrium constants *in vivo* and *in vitro*, which we summarize in Table I. Although these values correspond to specific examples of *E. coli* promoters, like the $P_{lac}$ or the $P_{RM}$ promoters, we extend their reach by using them as "typical" parameters characteristic of bacterial promoters, with the idea being that we are trying to demonstrate the classes of effects that can be expected, rather than dissecting in detail any particular promoter. The rate of association for transcription factors to operators *in vivo* is assumed to be the same as the recently measured value for the Lac repressor, which is close to the diffusion limited rate [49].

Operator strength reflects how tightly operators bind their transcription factors, and it is quantitatively characterized by the equilibrium dissociation constant $K_{O-TF}$. The dissociation constant has units of concentration and is equal to the transcription factor concentration at which the probability for the operator to be occupied is 1/2. $K_{O-TF}$ is related to the association and dissociation rates by $K_{O-TF} = k_{off}/k_{on}^0$, where $k_{off}$ is the rate at which a transcription factor dissociates from the promoter, and $k_{on}^0$ is the association rate per molar of transcription factor (i.e. $k_{on} = k_{on}^0 [N_{TF}]$). For simplicity, we work on the assumption that the binding reaction is diffusion limited, $k_{on}^0$ is already close to its maximum possible value, so the only parameter that can differ from operator to operator is the dissociation rate: strong operators have slow dissociation rates, and weak operators have large dissociation rates.

Throughout this paper, we also make the assumption that the mean expression level is controlled by varying the intracellular concentration of transcription factors, a scenario that is very common experimentally [50,51,52]. We also assume that changing the intracellular concentration of transcription factors only affects the association rate of transcription factors to the operators, but the dissociation rate and the rates of transcription at each promoter state are not affected. In other words, $k_{off}$ is a constant parameter for each operator, and it is not changed when we change the mean by titrating the intracellular repressor level.

All of these general assumptions need to be revisited when studying a specific gene-regulatory system. Here our focus is on illustrating the general principles associated with different promoter architectures typical of those found in prokaryotes.

*Simulations*

To generate mRNA time traces, we applied the Gillespie algorithm [53] to the master equation described in the text. A single time step of the simulation is performed as follows: one of the set of possible trajectories is chosen according to its relative weight, and the state of the system is updated appropriately. At the same time, the time elapsed since the last step is chosen from an exponential distribution, whose rate parameter equals the sum of rate parameters of all possible trajectories. This process is repeated iteratively to generate trajectories that exactly reflect dynamics of the underlying master equation. For the figures, simulation lengths were set long enough for the system to reach steady state and for a few promoter state transitions to occur.

To generate the probability distributions, it is convenient to reformulate the entire system of mRNA master equations in terms of a single matrix equation. To do this, we first define a vector

$$P = \begin{bmatrix} p_0(0) \\ p_1(0) \\ \vdots \\ p_N(0) \\ p_0(1) \\ \vdots \\ p_N(1) \\ p_0(2) \\ \vdots \end{bmatrix} = \begin{bmatrix} \vec{p}(0) \\ \vec{p}(1) \\ \vec{p}(2) \\ \vdots \end{bmatrix}$$

(16)

where $p_i(m)$ is the joint probability of having $m$ mRNA while in the $i$th promoter state. Then the master equation for time evolution of this probability vector is

$$\frac{d\mathbf{P}}{dt} = \begin{vmatrix} \hat{K} - \hat{R} & \hat{\Gamma} & 0 & 0 & \cdots \\ \hat{R} & \hat{K} - (\hat{R} + \hat{\Gamma}) & 2\hat{\Gamma} & 0 & \\ 0 & \hat{R} & \hat{K} - (\hat{R} + 2\hat{\Gamma}) & 3\hat{\Gamma} & \\ 0 & 0 & \hat{R} & \hat{K} - (\hat{R} + 3\hat{\Gamma}) & \\ \vdots & & & & \ddots \end{vmatrix} \bullet \mathbf{P}$$

(17)

where each element of the matrix is itself an N by N matrix as described in the text. Then finding the steady-state distribution $\mathbf{P}_{ss}$ is equivalent to finding the eigenvector of the above matrix associated with eigenvalue 0. To perform this calculation numerically, one must first choose an upper bound on mRNA copy number in order to work with finite matrices. In this work, we chose an upper bound six standard deviations above mean mRNA copy number as an initial guess, and then modified this bound if necessary. Computations were performed using the SciPy (Scientific Python) software package.

## RESULTS

### Single repression architecture: operator strength

We first investigate a promoter architecture consisting of a single repressor binding site, and examine how operator strength affects intrinsic variability in gene expression. Although this particular mode of gene regulation has been well studied theoretically before [2,20,36,37,45], it is a useful starting point in illustrating the utility of this class of models. Within this class of models, when the repressor is bound to the operator, it interferes with transcription initiation and transcription does not occur. When the repressor dissociates and the operator is free, RNAP can bind and initiate transcription at a constant rate $r$. The probability per unit time that a bound repressor dissociates is $k_R^{off}$, and the probability per unit time that a free repressor binds the empty operator is $k_R^{on} = k_{on}^0 [N_R]$, where $k_{on}^0$ is the second-order association constant and $[N_R]$ is the intracellular repressor concentration. The rate of mRNA degradation per molecule is $\gamma$. This mechanism is illustrated in Figure 2A.

We compute the mean and the Fano factor for this architecture following the algorithm described in the Mathematical Methods section. The kinetic rate and transcription rate matrices $\hat{K}$ and $\hat{R}$ are shown in Table SI. For this simple architecture, the mean of the mRNA probability distribution and the Fano factor take simple analytical forms:

$$\langle m \rangle = \frac{r}{\gamma} \frac{k_R^{off}}{k_R^{off} + k_R^{on}} = \frac{r}{\gamma} \frac{1}{1 + k_R^{on}/k_R^{off}}, \tag{18}$$

$$\eta^2 = \frac{1}{\langle m \rangle} + \frac{k_R^{on}}{k_R^{off}} \frac{\gamma}{\gamma + k_R^{off} + k_R^{on}}. \tag{19}$$

Using the relationship between $k_R^{on}$ and the intracellular concentration of repressor, we can write the mean as:

$$\langle m \rangle = \frac{r}{\gamma} \frac{1}{1 + k_{on}^0 [N_R]/k_R^{off}} = \langle m \rangle_{max} \frac{1}{1 + [N_R]/K_{OR}}. \tag{20}$$

Here we have defined the equilibrium dissociation constant between the repressor and the operator as: $K_{OR} = k_R^{off}/k_{on}^0$. It is interesting to note that equation (20) could have been derived using the thermodynamic model approach [33,34,41,54]. In particular we see that this expression is equal to the product of the maximal activity in the absence of repressor $\langle m \rangle_{max} = r/\gamma$, and the so-called fold-change in gene expression: $(1 + k_R^{on}/k_R^{off})^{-1} = (1 + [N_R]/K_{OR})^{-1}$ [34].

The Fano factor for the mRNA distribution can be computed from equation (12) and we obtain:

$$Fano = 1 + \frac{k_R^{on}}{\left(k_R^{off} + k_R^{on}\right)} \frac{r}{\left(\gamma + k_R^{off} + k_R^{on}\right)}, \qquad (21)$$

which is also shown as the first entry of Table II. In many experiments [4,6,9,51], the concentration of repressor inside the cell $[N_R]$ (and therefore the association rate $k_R^{on} = k_{on}^0 [N_R]$) can be varied by either expressing the repressor from an inducible promoter, or by adding an inducer that binds directly to the repressor rendering it incapable of repressing transcription. When such an operation is performed, the only parameter that is varied is typically $k_R^{on}$, and all other kinetic rates are constant. The Fano factor can thus be re-written as a function of the mean mRNA, and we find:

$$Fano = 1 + \langle m \rangle \left( \frac{1 - \langle m \rangle/\langle m \rangle_{max}}{k_R^{off}/\gamma + \langle m \rangle/\langle m \rangle_{max}} \right). \qquad (22)$$

Therefore, for any given value of the mean, the Fano factor depends only on two parameters: the maximal mRNA or protein expression per cell, and a parameter that reflects the strength of binding between the repressor and the operator: $k_R^{off}/\gamma$. Equations (20) and (21) reveal that changes in the mean due to repressor titration affect the noise as well as the mean. Since neither the repressor dissociation rate $k_R^{off}$ nor the mRNA degradation rates are affected by the concentration of repressors, $k_R^{off}/\gamma$ is a constant parameter that will determine how large the cell-to-cell variability is: The Fano factor is maximal for promoters with very strong operators, ($k_R^{off} \ll \gamma$), and it goes to 1 (i.e., the distribution tends to a Poisson distribution) when the operator is very weak and the rate of dissociation extremely fast ($k_R^{off} \gg \gamma$). In the latter limit of fast promoter kinetics, the fast fluctuations in promoter occupancy are filtered by the long lifetime of mRNA. Effectively, mRNA degradation acts as a low-pass frequency filter [55,56], and fast fluctuations in promoter occupancy are not propagated into mRNA fluctuations. Therefore, promoters with strong operators are expected to be noisier than promoters with weak operators [57].

This effect is illustrated in Figures 2B and 2C, where we show the normalized variance and the Fano factor, as a function of the fold-change in the mean mRNA concentration for a single strong operator whose dissociation rate is $k_R^{off} = 0.0027 \text{s}^{-1}$ (a value that is representative of well characterized repressor-operator interactions such as the Lac repressor-*lac*O1, or the cI2-$\lambda$ $O_{R1}$), and for a single operator whose dissociation rate $k_R^{off}$ is 10 times faster. The Poisson noise is shown for reference. The level of variability is always smaller for the weak operator than for the strong operator, due to faster promoter switching leading to smaller mRNA fluctuations and a more Poisson-like mRNA distribution (Figure 2E), in which most cells are close to the mean. Slow dissociation, on the other hand, causes slower promoter fluctuations and highly non Poissonian mRNA distributions, with few cells near the mean expression level (see Figure 2E, strong promoter). In Figure 2D we plot the fold-change in protein noise due to gene regulation for the simple repression architecture. As expected, we find that the effect of operator strength in protein noise is qualitatively identical to what we found for mRNA. Since the same can be said of all the rest of architectures studied, we will limit the discussion to mRNA noise for the rest of the architectures, with the understanding that for the class of models considered here, all the conclusions about the effect of promoter architecture in cell-to-cell variability that are valid for mRNA, are true for protein noise as well.

An example of this first architecture is a simplified version of the $P_{lacUV5}$ promoter, which consists of one single operator overlapping with the promoter. Based on a simple kinetic model of repression, in which the Lac repressor competes with RNAP for binding at the promoter, we can write down the $\hat{K}$ and $\hat{R}$ matrices and compute the cell-to-cell variability in mRNA copy number. The matrices are presented in Table SI. Based on our previous analysis, we know that stronger operators are expected to cause larger noise and higher values of the Fano factor than weaker operators. Therefore, we expect that if we replace the wild-type O1 operator by the 10 times weaker O2 operator, or by the ~500 times weaker operator O3, the fold-change in noise should go down. Entering our best estimates and measurements for the kinetic parameters involved, we find that noise is indeed much larger for O1 than for O2, and it is negligible for O3. This prediction is presented as an inset in Figure 2C.

## Promoters with two repressor-binding operators

Dual repression occurs when promoters contain two or more repressor binding sites. Here, we consider three different scenarios for architectures with two operators: 1) repressors bind independently to the two operators, 2) repressors bind cooperatively to the two operators and 3) one single repressor may be bound to the two operators simultaneously, by looping DNA. At the molecular level, cooperative repression is achieved by two weak operators that form long-lived repressor-bound complexes when both operators are simultaneously occupied. Transcription factors may stabilize each other either through direct protein-protein interactions [54], or through indirect mechanisms mediated by alteration of DNA conformation [58].

**Cooperative and Independent repression**

The kinetic mechanisms of gene repression for both the cooperative and independent repressor architectures are reproduced in Figure 3A. For simplicity, we assume that both sites are of equal strength, so the rates of association and dissociation to both sites are equal. Cooperative binding is reflected in the fact that the rate of dissociation from the state where the two operators are occupied is slower (by a factor $\omega \ll 1$) than the dissociation from a single operator. A typical value of ω for cooperative binding is on the order of $\sim 10^{-3} - 10^{-2}$ [51,54]. By way of contrast, independent binding is characterized by a value of $\omega = 1$, which reflects the fact that the rate of dissociation from each operator is not affected by the presence of the other operator.

The $\hat{K}$ and $\hat{R}$ matrices for these two architectures are defined in Table SI. Using these matrices, we can compute the mean gene expression and the Fano factor for these two architectures as a function of the concentrations of repressor. The resulting expression for the fold-change in noise is shown as entry number 3 of Table II. As shown in Figure 3B, the noise for cooperative repression is substantially larger than for the independent repression architecture. The high levels of intrinsic noise associated with cooperative repression can be understood intuitively in terms of the kinetics of repressor-operator interactions. The lifetime of the states where only one repressor is bound to either one of the two weak operators is so short that only rarely another repressor will bind to the second operator before the first repressor dissociates. This makes simultaneous binding of two repressors to the two operators a rare event. However, when it occurs, the two repressors stabilize each other, forming a very long-lived complex with the operator DNA. This mode of repression, with rare but long-lived repression events, is intrinsically very noisy, since the promoter switches slowly between active (unrepressed) and inactive (repressed) states, generating wide bimodal distributions of mRNA (see figure 3C). On the other hand, independent binding to two operators causes more frequent transitions between repressed and unrepressed states, leading to lower levels of intrinsic noise and unimodal mRNA distributions (see Figure 3C).

As an example of this architecture, we consider a simplified version of the lytic phage-$\lambda$ $P_R$ promoter, which is controlled by the $\lambda$ phage lysogenic repressor cI. The wild-type $P_R$ promoter consists of three proximal repressor binding sites, $O_{R1}$, $O_{R2}$ and $O_{R3}$, with different affinities for the repressor ($O_{R2}$ is ~25 times weaker than $O_{R1}$) [59], and three distal operators $O_{L1}$, $O_{L2}$ and $O_{L3}$. For simplicity, we consider a simpler version of $P_R$, harboring a deletion of the three distal operators. In the absence of these operators, the $O_{R3}$ operator plays only a very minor role in the repression of this promoter, and it can be ignored [51,60]. We are then left with only $O_{R1}$ and $O_{R2}$. The cI repressor binds cooperatively to $O_{R1}$ and $O_{R2}$, and that cooperativity is mediated by direct protein-protein interactions between cI bound at each operator [60]. Mutant forms of cI that are cooperativity deficient (i.e., not able to bind cooperatively to the promoter) have been designed [61]. We can use the stochastic model of gene regulation described in the theory section, to make precise predictions that address the issue of how cooperativity affects stochastic gene regulation. In the inset in Figure 3B, we compare the normalized variance of the mRNA distribution, both for wild-type cI repressor, and for a cooperativity deficient mutant such as Y210H [61]. The cooperative repressor is predicted to have significantly larger promoter noise than the cooperativity deficient mutant.

**Simultaneous binding of one repressor to two operators: DNA Looping**

Repression may also be enhanced by the presence of distant operators, which stabilize the repressed state by allowing certain repressors to simultaneously bind to both distant and proximal operators, forming a DNA loop [62,63]. The $P_{lac}$ promoter is a prominent example of this architecture. The kinetic mechanism of repression characterizing this promoter architecture is presented in Figure 4A. The repressor only prevents transcription when it is bound to the main operator Om, but not when it is only bound to the auxiliary operator Oa. DNA loop formation is characterized by a kinetic rate $k_{loop} = k_{on}^0 [J]$ where $[J]$, the looping J-factor, can be thought of as the local concentration of repressor in the vicinity of one operator when the repressor is bound to the other operator [33,34]. The rate of dissociation of the operator-repressor complex in the looped conformation is given by $k_{unloop} = c\, k_R^{off}$. The parameters $[J]$ and $c$ have both been measured *in vitro* for the particular case of the Lac repressor [42,64], and also estimated from *in vivo* data [33,65]. The $\hat{K}$ and $\hat{R}$ matrices for this architecture are defined in Table SI. We use these matrices to compute the mean and the noise strength, according to equations (5-12) resulting in the 5$^{th}$ entry of Table II.

We first examine how the presence of the auxiliary operator affects the level of cell-to-cell variability in mRNA expression. In Figure 4B we compare the Fano factor in the absence of the auxiliary operator with the Fano factor in the presence of the auxiliary operator, which is assumed to be of the same strength as the main operator. We use parameters in Table I, and we first assume that the dissociation rate of the operator-repressor complex in the looped state is the same as the dissociation rate in the unlooped state, so $c = 1$ and $k_{unloop} = k_R^{off}$. This assumption is supported by single-molecule experiments in which the two operators are on the same side of the DNA double-helix, separated by multiples of the helical period of DNA [42,64]. Under these conditions we find that the presence of an auxiliary operator results in a larger Fano factor, in spite of the fact that the auxiliary operator *Oa* does not stabilize the binding of the repressor to the main operator *Om*. Interestingly, we find that the Fano factor is maximal at intermediate concentrations of repressor for which only one repressor is bound to the promoter, making the simultaneous occupancy of the auxiliary and main operators mediated by DNA looping possible. In contrast, the Fano factor is identical to that of the simple repression case if the concentration of repressor is so large that it saturates both operators and looping never occurs. It had been previously hypothesized that DNA looping might be a means to reduce noise in gene expression, due to rapid re-association kinetics between *Om* and a repressor that is still bound to *Oa*, which may cause short and frequent bursts of transcription [65,66]. Here, by applying a simple stochastic model of gene regulation, we show that the presence of the auxiliary operator does not, by itself, decrease cell-to-cell variability. On the contrary, it is expected to increase it. The reason for this increase is that the rate of dissociation from the main operator is not made faster by DNA looping; instead the presence of the auxiliary operator causes the repressor to rapidly rebind the main operator, extending the effective period of time when the promoter is repressed.

Indeed, we find that only if the dissociation rate for a repressor in the looped state is much faster than in the unlooped state, the presence of the auxiliary operator might reduce the cell-to-cell variability. To illustrate this limit, we have assumed a value of $c = 100$, so that $k_{unloop} = 100\ k_R^{off}$, and find that the Fano factor goes down, below the expectation for the simple repression architecture. A modest increase in the dissociation rate in the looped conformation has been reported in recent single-molecule experiments for promoter architectures in which the two operators are out of phase (located on different faces of the DNA) [42].

An example of this type of architecture is a simplified variant of the $P_{lacUV5}$ promoter, which consists of one main operator and one auxiliary operator upstream from the promoter. The kinetic mechanism of repression is believed to be identical to the one depicted in Figure 4A [5,42,64,65]. We can use the stochastic model of gene regulation described in the theory section to make precise predictions that will test this kinetic model of gene regulation by DNA looping. We find that the kinetic model predicts that, if we move the center of the auxiliary operator further upstream from its wild-type location, in increments of distance given by the helical period of the DNA, such that both operators stay in phase, the fold-change in noise should behave as represented in Figure 4D, getting closer to the fold-change in noise corresponding to having only the main operator as *Oa* is moved away from the promoter.

## Simple Activation

Transcriptional activators bind to specific sites at the promoter from which they increase the rate of transcription initiation by either direct contact with one or more RNAP subunits or indirectly by modifying the conformation of DNA around the promoter [58]. The simplest example of an activating promoter architecture consists of a single binding site for an activator in the vicinity of the RNAP binding site. When the activator is not bound, transcription occurs at a low basal rate. When the activator is bound, transcription occurs at a higher, activated rate. Stochastic association and dissociation of the activator causes fluctuations in transcription rate which in turn cause fluctuations in mRNA copy number.

This simple activation architecture was illustrated in Figure 1A. The $\hat{K}$ and $\hat{R}$ matrices for this architecture are given in Table SI. Solving equations (5-8) for this particular case, we find that the mean mRNA per cell for this simple mechanism takes the form:

$$\langle m \rangle = \frac{r_2}{\gamma} \frac{k_A^{on}}{k_A^{on} + k_A^{off}} + \frac{r_1}{\gamma} \frac{k_A^{off}}{k_A^{on} + k_A^{off}} \qquad (23)$$

The mean mRNA can be changed by adjusting the intracellular concentration of the activator. The rate at which one of the activators binds to the promoter is proportional to the activator concentration: $k_A^{on} = k_{on}^0 [N_A]$. Following the same argument as we used in the simple repression case, the equilibrium dissociation constant for the activator-promoter interaction is given by $K_{OA} = k_A^{off}/k_{on}^0$. Finally, it is convenient to define the enhancement factor: the ratio between the rate of transcription in the active and the basal states $f = r_2/r_1$. The mean mRNA can be written in terms of these parameters as:

$$\langle m \rangle = \frac{r_1}{\gamma} \left( \frac{K_{OA}}{[N_A] + K_{OA}} + f \frac{[N_A]}{[N_A] + K_{OA}} \right). \qquad (24)$$

The Fano factor can be computed using equations (5-12) and it is shown as entry 2 of Table II. By writing $k_A^{on}$ as a function of the mean:

$$Fano = 1 + \langle m \rangle \left( \frac{f - \langle m \rangle/\langle m \rangle_{basal}}{\langle m \rangle/\langle m \rangle_{basal}} \right)^2 \frac{\langle m \rangle/\langle m \rangle_{basal} - 1}{\left(f - \langle m \rangle/\langle m \rangle_{basal}\right) + \frac{k_A^{off}}{\gamma}(f - 1)} \qquad (25)$$

With these equations in hand, we explore how operator strength affects noise in gene expression in the case of activation. Stronger operators bind to the activator more tightly than weak operators, leading to longer residence times of the promoter in the active state.

In Figure 5A we plot the Fano factor as a function of the fold-change in mean expression for a strong operator as well as a 10 times weaker operator. We have used the parameters in Table I. Just as we saw for the simple repression architecture, it is also true for the simple activation architecture that stronger operators cause larger levels of noise for activators than weaker operators.

To get a sense of the differences between these two standard regulatory mechanisms, we compare simple repression with simple activation. In Figure 5B, we plot the Fano factor as a function of the mean for a repressor and an activator with identical dissociation rates. We assume that the promoter switches between a transcription rate $r = 0$ in its inactive state (which happens when the repressor is bound in the simple repression case, or the activator is not bound in the simple activation case), and a rate equal to $r = 0.33 mRNA\ s^{-1}$ (see Table I) in the active state (repressor not bound in the simple repression case, activator bound in the simple activation case). As shown in Figure 5B, at low expression levels the simple activation is considerably (>20 times) noisier than the simple repression promoter. At high expression levels both architectures yield very similar noise levels. A low level of gene expression may be achieved either by low concentrations of an activator, or by high concentrations of a repressor. Low concentrations of an activator will lead to rare activation events. High concentrations of a repressor will lead to frequent but short-lasting windows of time for which the promoter is available for transcription. As a result, and as we illustrate in Figure 5C, the activation mechanism leads to bursty mRNA expression whereas the repressor leads to Poissonian mRNA production. This result suggests that in order to maintain a homogeneously low expression level, a repressive strategy in which a high concentration of repressor ensures low expression levels may be more adequate than a low activation strategy.

An example of simple activation is the wild-type $P_{lac}$ promoter, which is activated by CRP when complexed with cyclic AMP (cAMP). CRP is a ubiquitous transcription factor, and is involved in the regulation of dozens of promoters, which contain CRP binding sites of different strengths [67]. In the inset of Figure 5A we include CRP as an example of simple activation, and make predictions for how changing the wild-type CRP binding site in the $P_{lac}$ promoter by the CRP binding site of the $P_{gal}$ promoter (which is ~8 times weaker [68]), should affect the Fano factor. As expected from our analysis of this class of promoters, the noise goes down.

## Dual Activation: independent and cooperative activation

Dual activation architectures have two operator binding sites. Simultaneous binding of two activators to the two operators may lead to a larger promoter activity in different ways. For instance, in some promoters each of the activators may independently contact the polymerase, recruiting it to the promoter. As a result, the probability to find RNAP bound at the promoter increases and so does the rate of transcription [33,69]. In other instances, there is no increase in enhancement factor when the two activators are bound. However, the first activator recruits the second one through protein-protein or protein-DNA interactions, stabilizing the active state and increasing the fraction of time that the promoter is active [60]. These two modes are not mutually exclusive, and some promoters exhibit a combination of both mechanisms [70].

We first investigate the effect of dual activation in the limit where binding of the two transcription factors is not cooperative, and activators bound at the two operators may independently recruit the polymerase, and compare this architecture with a simple activation architecture. The mechanism of activation is depicted in Figure 6A, and matrices $\hat{K}$ and $\hat{R}$ are presented in Table SI. For simplicity, we assume that both operators have the same strength, and both have the same enhancement factor $f = r_2/r_1 = r_3/r_1$. When the two activators are bound, the total enhancement factor is given by the product of the individual enhancement factors, which in this case is $f \times f = r_4/r_1$ [33]. All of the other relevant kinetic parameters are given in Table I. The Fano factor is plotted in figure 6B. We find that compared to the simple operator architecture, the second operator increases the level of variability, even when binding to the operators is non-cooperative.

We then ask whether this is also true when the binding of activators is cooperative. We assume a small cooperativity factor of $\omega = 0.1$. Just as we found for repressors, cooperative binding of activators generates larger cell-to-cell variability than independent binding, which in turn generates larger cell-to-cell variability than simple activation. This is illustrated in the stochastic simulation in figure 6C. As expected the dual activation architectures are noisier than the simple activation, presenting rare but long lived activation events that lead to large fluctuations in mRNA levels. In contrast, simple activation architecture leads to more frequent but less intense activation events.

Together with the results from the dual repressor mechanism, these results indicate that multiplicity in operator number may introduce significant intrinsic noise in gene expression and be a source of cell-to-cell variability in gene expression. Operators commonly appear in multiple repeats in eukaryotic promoters [2,71,72], and operator repeats are often found in prokaryotic promoters too [60,69,73]. When transcription factors bind to these operator repeats in a cooperative manner, the sensitivity of the fold-change in gene expression to changes in transcription factor concentration will be larger [33], and this increase in sensitivity may be accompanied by large levels of cell-to-cell variability.

An example of cooperative activation is the lysogenic phage-$\lambda$ $P_{RM}$ promoter [60]. This promoter contains three operators ($O_{R1}, O_{R2}$ and $O_{R3}$) for the cI protein, which acts as an activator. When $O_{R2}$ is occupied, cI activates transcription. $O_{R1}$ has no direct effect on the transcription rate, but it helps recruit cI to $O_{R2}$, since cI binds cooperatively to the two operators. Finally, $O_{R3}$ binds cI very weakly, but when it is occupied, $P_{RM}$ becomes repressed. There are variants of this promoter [51] that harbor mutations in $O_{R3}$ that make it unable to bind cI. In Figure 6D, we include one of these variants, $r1-P_{RM}$ [51] as an example of dual activation, and we present a theoretical prediction for the promoter noise as a function of the mean mRNA. We examine the role of cooperativity by comparing the wild-type cI, with a cooperativity deficient mutant. We find that the cooperative activator causes substantially larger cell-to-cell variability than the mutant, emphasizing our expectation that cooperativity may cause substantial noise in gene expression in bacterial promoters such as $P_{RM}$.

## DISCUSSION AND CONCLUSIONS

The DNA sequence of a promoter encodes the binding sites for transcriptional regulators. In turn, the collection of these regulatory sites, known as the architecture of the promoter, determines the mechanism of gene regulation. The mechanism of gene regulation, determines the transcriptional response of a promoter to a specific input, in the form of the concentration of one or more transcription factors or inducer molecules. In recent years we have witnessed an increasing call for quantitative models of gene regulation that can serve as a conceptual framework for reflecting on the explosion of recent quantitative data, testing hypotheses, and proposing new rounds of experiments [34,74,75]. Much of this data has come from bulk transcription experiments with large numbers of cells, in which the average transcriptional response from a population of cells (typically in the form of the level of expression of a reporter protein) was measured as a function of the concentration of a transcription factor or inducer molecule [51,76]. Thermodynamic models [34,41,54] of gene regulation are a general framework for modeling gene regulation and dealing with this kind of bulk transcriptional regulation experiments. This class of models has proven to be very successful at predicting gene expression patterns from the promoter architecture encoded in the DNA sequence [50,74,75,76,77,78]. However, a new generation of experiments now provides information about gene expression at the level of single-cells, with single-molecule resolution [5,6,7,9,10,11,14,15,47,52]. These experiments provide much richer information than just how the mean expression changes as a function of an input signal: it tells us how that response is spread among the population of cells, distinguishing homogeneous responses, in which all cells express the same amount of proteins or mRNA for the same input, from heterogeneous responses in which some cells achieve very high expression levels while others maintain low expression. Thermodynamic models are unable to explain the single-cell statistics of gene expression, and therefore are an incomplete framework for modeling gene regulation at the single-cell level.

A class of stochastic kinetic models have been formulated that make it possible to calculate either the probability distribution of mRNA or proteins per cell or its moments, for simple models of gene regulation involving one active and one inactive promoter state [36,37,45,79]. Recently, we have extended that formalism to account for any number of promoter states[32], allowing us to model any promoter architecture within the same mathematical framework.

Armed with this model, we can now ask how promoter architecture affects not only the response function, but also how that response is distributed among different cells.

In this paper we have explored the feasibility of this stochastic analog of thermodynamic models as a general framework to understand gene regulation at the single-cell level. Using this approach we have examined a series of common promoter architectures of increasing complexity, and established how they affect the level of cell-to-cell variability of the number of mRNA molecules, and proteins, in steady state. We have found that the level of variability in gene expression for bacterial promoters is expected to be larger than the simple Poissonian expectation, particularly for mRNA and short-lived proteins. We have investigated how the level of variability generated by a simple promoter consisting of one single operator differs from more complex promoters containing more than one operator, and found that the presence of multiple operators increases the level of cell-to-cell variability even in the absence of cooperative binding. Cooperative binding makes the effect of operator multiplicity even larger. We also found that operator strength is one of the major determinants of cell-to-cell variability. Strong operators cause larger levels of cell-to-cell variability than weak operators. We have also examined the case where one single repressor may bind simultaneously to two operators by looping the DNA in between. We have found that the stability of the DNA loop is the key parameter in determining whether DNA looping increases or decreases the level of variability, suggesting a potential role of DNA mechanics in regulating cell-to-cell variability.

We have examined the difference between activators and repressors, and found that repressors generate less cell-to-cell variability than activators at low expression levels, whereas at high expression levels repressors and activators generate similar levels of cell-to-cell variability. We conclude that induction of gene expression by increasing the concentration of an activator leads to a more heterogeneous response at low and moderate expression levels than induction of gene expression by degradation, sequestration or dilution of a repressor. In addition, we have used this model to make quantitative predictions for a few well characterized bacterial promoters, connecting the kinetic mechanism of gene regulation that we believe applies for these promoters *in vivo* with single-cell gene expression data. Direct comparison between the model and experimental data offers an opportunity to validate these kinetic mechanisms of gene regulation.

*Comparison with experimental results*

Though there are several examples of experiments in which perturbations to the promoter architecture have been observed to affect the cell-to-cell variability in gene expression, only a few of these measured intrinsic noise, the magnitude that is directly comparable with the model in this paper. The two experiments in which the intrinsic mRNA noise was measured (both were done in yeast cells), report that operator multiplicity leads to higher noise levels [2,6], which is in agreement with our analysis. In a separate study, Murphy and co-workers [16] also found that the multiplicity of Tet repressor binding sites leads to larger levels of cell-to-cell variability. However, this study measured the total noise strength, and did not isolate the intrinsic noise, so the observed increase in noise strength may have other origins. Another recent article by the same group [3] reports an overall decrease in cell-to-cell variability in protein expression by mutations that weaken the affinity of a transcription factor binding site (TATA box) for its transcription factor (TATA box binding protein). This is also in agreement with the expectation from the model. However, this experiment also measured total noise levels, rather than intrinsic noise. Finally, the effect of DNA looping on the total noise strength has also been recently measured for the $P_{lacUV5}$ promoter in *E. coli* [5]. Using a novel single-protein counting technique, Choi and co-workers measured protein distributions for promoters whose auxiliary operator had been deleted (leaving them with a simple repression architecture), and compared them to promoters with the auxiliary operator O3 present, which allows for DNA looping. They report a reduction in protein noise due to the presence of O3, which according to our analysis, may indicate that the dissociation of the repressor from the looped state is faster than the normal dissociation rate. However, their measurements also reflect the total noise, and not only the intrinsic part, so the explanation may lie elsewhere. These results emphasize the need for more experiments in which the intrinsic noise is isolated and measured directly.

More recently, several impressive experimental studies have measured the noise in mRNA in bacteria for a host of different promoters ([80], Ido Golding, private communication). In both of these cases, simplified low-dimensional models which do not attend to the details of the promoter architecture have been exploited to provide a theoretical framework for thinking about the data. Our own studies indicate that the differences between a generic two-state model and specific models that attempt to capture the details of a given architecture are sometimes subtle and that the acid test of ideas like those presented in this paper can only come from experiments which systematically tune parameters such as the repressor concentration for a given transcriptional architecture.

*Future Directions*

Some recent theoretical work has analyzed the effect of cooperative binding of activators in the context of particular examples of eukaryotic promoters [81,82]. The main focus of this study are bacterial promoters. The simplicity of the microscopic mechanisms of transcriptional regulation for bacterial promoters makes them a better starting point for a systematic study like the one we propose. However, many examples of eukaryotic promoters have been found whose architecture affects the cell-to-cell variability [1,2,3,6,16]. Although the molecular mechanisms of gene regulation in these promoters are much more complex, with many intervening global and specific regulators [83], the stochastic model employed in this paper can be applied to any number of promoter states, and thus can be applied to these more complex promoters. Recent experimental work is starting to reveal the dynamics of nucleosomes and transcription factors with single-molecule sensitivity [84,85], allowing the formulation of quantitative kinetic and thermodynamic mechanistic models of transcriptional regulation at the molecular level [74,78]. The framework for analyzing gene expression at the single-cell level developed in this paper will be helpful to investigate the kinetic mechanisms of gene regulation in eukaryotic promoters, as the experimental studies switch from ensemble, to single-cell.

*Shortcomings of the approach*

Although the model of transcriptional regulation used in this paper is standard in the field, it is important to remark that it is a very simplified model of what really happens during transcription initiation. There are many ways in which this kind of model can fail to describe real situations. For instance, mRNA degradation requires the action of RNases. These may become saturated if the global transcriptional activity is very large, and degradation becomes non-linear [56]. Transcription initiation and elongation are assumed to be jointly captured in a single constant rate of mRNA synthesis for each promoter state. This is an oversimplification also. When considered explicitly, and in certain parameter ranges, the kinetics of RNAP-promoter interaction may cause noticeable effects in the overall variability [46]. Similarly, as pointed out elsewhere [86,87,88], translational pausing, backtracking or roadblocking may also cause significant deviations in mRNA variability from the predictions of the model used in this paper. How serious these deviations are depends on the specifics of each promoter-gene system. The model explored in this paper here also assumes that the cell is a well-mixed environment. Deviations from that approximation can significantly affect cell-to-cell variability [57,89]. Another simplification refers to cell growth and division, which are not treated explicitly by the model used in this paper: cell division and DNA replication cause doubling of gene and promoter copy number every cell cycle, as well as binomial partition of mRNAs between mother and daughter cells [8]. In eukaryotes, mRNA often needs to be further processed by the splicing apparatus before it becomes transcriptionally active. It also needs to be exported out of the nucleus, where it can be translated by ribosomes.

To study the effect of transcription factor dynamics on mRNA noise we assume that the unregulated promoter produces mRNA in a Poisson manner, at a constant rate. This assumption can turn out to be wrong if there is another process, independent of transcription factors, that independently turns the promoter on and off. In eukaryotes examples of such processes are nucleosome positioning and chromatin remodeling, while in prokaryotes similar processes are not as established but could include the action of non-specifically bound nucleoid proteins such as HU and HNS, or DNA supercoiling. Experiments that measure cell-to-cell distributions of mRNA copy number in the absence of transcription factors (say without Lac repressor for the lac operon case) can settle this question. In case the Fano factor (variance divided by the mean) for this distribution is not one (as expected for a Poisson distribution) this can signal a possible transcription factor-independent source of variability. The stochastic models studied here can be extended to account for this situation. For example the promoter can be made to switch between an on and an off state, where the transcription factors are allowed to interact with promoter DNA only while it is in the on state. In this case the mRNA fluctuations produced by an unregulated promoter will not be Poissonian. One can still investigate the affect of transcription factors by measuring how they change the nature of mRNA fluctuations from this new base-line. Comparison of this extended model with single-cell transcription experiments would then have the exciting potential for uncovering novel modes of transctriptional regulation in prokaryotes.

We have also assumed that when transcription factors dissociate from the operator, they dissociate into an averaged out, well-mixed, mean-field concentration of transcription factors inside the cell. The possibility of transcription factors being recaptured by the same or another operator in the promoter right after they fall off the operator is not captured by the class of models considered here. Recent *in vivo* experiments suggest that this scenario may be important in yeast promoters containing arrays of operators [6].

In spite of all of the simplifications inherent in the class of models analyzed in this paper, we believe they are an adequate jumping off point for developing an intuition about how promoter architecture contributes to variability in gene expression. Our approach is to take a highly simplified model of stochastic gene expression, based on a kinetic model for the processes of the central dogma of molecular biology, and add promoter dynamics explicitly to see how different architectural features affect variability. This allows us to isolate the effect of promoter dynamics, and develop an intuitive understanding of how they affect the statistics of gene expression.

It must be emphasized, however, that the predictions made by the model may be wrong if any of the complications mentioned above are significant. This is not necessarily a bad outcome. If the comparison between experimental data and the predictions made by the theory for any particular system reveals inconsistencies, then the model will need to be refined and new experiments are required to identify which of the sources of variability that are not accounted for by the model are in play. In other words, experiments that test the quantitative predictions outlined stand a chance of gaining new insights about the physical mechanisms that underlie prokaryotic transcriptional regulation.

## ACKNOWLEDGEMENTS

We wish to thank Tom Kuhlman, Ido Golding, Terry Hwa, Bob Schleif, Paul Wiggins, Felix Hol, Narendra Mahreshri, T.L. To, C.J. Zopf and Jeff Gelles for many helpful discussions. AS wishes to thank Jeff Gelles and all members of the Gelles lab for their continuous support, as well as Melisa Osborne for her help in editing the manuscript. RP, HGG and DJ were supported by NIH grants R01 GM085286-01S and R01 GM085286 and National Institutes of Health Director's Pioneer Award grant DP1 OD000217. JK acknowledges support from National Science Foundation grant DMR-0706458. AS was supported by grants GM81648 and GM43369 from the National Institutes of Health

# FIGURE CAPTIONS

**Figure 1: Two-state promoter.** (A) Simple two-state bacterial promoter undergoing stochastic activation by a transcriptional activator binding to a single operator site. The rates of activator association and dissociation are given by $k_A^{on}$ and $k_A^{off}$, respectively and the rates of mRNA production for the basal and active states are $r_1$ and $r_2$ respectively. The mRNA degradation rate is assumed to be constant for each molecule, and is given by the parameter $\gamma$. (B) Trajectories and weights. List of all possible stochastic transitions affecting either the copy number of mRNA (m) or the state of the promoter. State 1 has the operator free. State 2 is the activator bound state. The weights represent the probability that each change of state will occur during a time increment $\Delta t$. The master equation is constructed based on these rules. (C) Cartoon representation of the kinetic rate matrix $\hat{K}$. The diagonal elements represent the net rate at which the promoter abandons each state. For instance, element $\{\hat{K}\}_{11}$ is the rate at which the promoter abandons state 1 due to stochastic association of the activator with the promoter: $\{\hat{K}\}_{11} = -k_A^{on}$, and element $\{\hat{K}\}_{22} = -k_A^{off}$ is the rate of dissociation of the activator from the promoter, abandoning state 2. The non-diagonal element $\{\hat{K}\}_{21} = k_A^{on}$ is the rate at which the promoter makes a transition from state 1 to state 2 (by dissociation association of one activator to the promoter), and the non-diagonal element $\{\hat{K}\}_{12} = k_A^{off}$ is the rate at which the promoter makes a transition from state 2 to state 1 (by dissociation of the activator). (D) The transcription rate matrix $\hat{R}$ contains, in its diagonal elements, the net rate of transcription at each promoter state. Element $\{\hat{R}\}_{11} = r_1$ is the rate of transcription in promoter state 1 and $\{\hat{R}\}_{22} = r_2$ is the rate of transcription in promoter state 2. (E) The vector $\vec{r} = (r_1, r_2)$ contains the rates of transcription at states 1 and 2, and is identical to the diagonal of matrix $\hat{R}$.

**Figure 2: Simple repression architecture.** (A) Kinetic mechanism of repression for an architecture involving a single repressor binding site. The repressor turns off the gene when it binds to the promoter (with rate $k_R^{on}$), and transcription occurs at a constant rate $r$ when the repressor falls off (with rate $k_R^{off}$). Promoter state 1 is defined as the state where repressor is not bound and the promoter is available for RNAP to bind and initiate transcription. State 2 is the state where the repressor is bound and RNAP is not able to initiate transcription. (B) Normalized variance as a function of the fold-change in mean mRNA. The parameters used are drawn from

Table I. The value of $k_R^{off} = 0.0023 s^{-1}$ from Table I corresponds to the *in vitro* dissociation constant of the Lac repressor Oid operator. The behavior for an off-rate 10-times higher is also plotted. As a reference for the size of the fluctuations, we show the normalized variance for a Poisson promoter. (C) Fano factor for two promoters bearing the same off-rates as in (B). Inset. Prediction for the Fano factor for the $\Delta_{O3}\Delta_{O2}P_{lacUV5}$ promoter, a variant of the $P_{lacUV5}$ promoter for which the two auxiliary operators are not present. The fold-change in mRNA noise is plotted as a function of the fold-change in mean mRNA expression for mutants of the promoter that replace O1 for Oid, O2 or O3. The parameters are taken from Table I and [33]. Lifetimes of the operator-repressor complex are 7min for Oid, 2.4min for O1, 11s for O2 and 0.47s for O3. (D) Fold-change in protein noise as a function of the fold-change in mean expression. As expected, the effect of operator strength is the same as observed for mRNA noise. (E) Time traces for promoter activity, mRNA and protein dynamics are shown for both the weak operator and the strong operator. The mRNA histograms are also shown. The weaker operator with a faster repressor dissociation rate leads to small promoter noise, and an mRNA probability distribution resembling a Poisson distribution, in which most cells express mRNA near the population average. In contrast, the stronger operator with a slower repressor dissociation rate, leads to larger promoter noise and strongly non-Poissonian mRNA statistics.

**Figure 3**: **Dual repression architecture.** (A) Kinetic mechanism of repression for a dual-repression architecture. The parameters $k_R^{off}$ and $k_R^{on}$ are the rates of repressor dissociation and association to the operators, and $\omega$ is a parameter reflecting the effect of cooperative binding in the dissociation rate. For independent binding, $\omega = 1$ and for cooperative binding $\omega = 0.013$ (see Table I). (B) Fold-change in the mRNA noise caused by gene regulation for independent and cooperative repression as a function of the mRNA concentration. The parameters used are shown in Table I. Inset: Prediction for a variant of the $\lambda P_R$ promoter where the upstream operators $O_{L1}, O_{L2}$ and $O_{L3}$ are deleted. The promoter mRNA noise is plotted as a function of the mean mRNA for both wild-type cI repressor (blue line) and a repressor mutant (Y210H) that abolishes cooperativity (red line). Parameters taken from [43,91]. The lifetime of the $O_{R1}$-cI complex is 4 min. Lifetime of $O_{R2}$-cI complex is 9.5s. (C) mRNA distribution for the same parameters used in (B).

**Figure 4: Repression by DNA looping.** (A) Kinetic mechanism of repression. $k_R^{off}$ and $k_R^{on}$ are the rates of repressor dissociation and association to the operators. The rate of loop formation is $k_{loop} = [J]k_R^0$, where $[J]$ can be thought of as the local concentration of repressor in the vicinity of operator when the repressor is bound to the other. The rate of dissociation of the operator-repressor complex in the looped conformation is given by $k_{unloop} = c\, k_R^{off}$. The parameter

$c$ captures the rate of repressor dissociation in the looped state relative to the rate of dissociation in a non-looped state. (B) Effect of DNA looping on cell-to-cell variability. The Fano factor is plotted as a function of the fold-change in the mean expression level, in the absence and presence of the auxiliary operator, and assuming that dissociation of the operator from $Om$ is the same in the looped and the unlooped state $c=1$. The presence of the auxiliary operator, which enables repression by DNA looping, increases the cell-to-cell variability. The regions over which the state with two repressors bound, the state with one repressor bound, or the looped DNA state are dominant are indicated by the shading in the background. The noise is larger at intermediate repression levels, where only one repressor is found bound to the promoter region, simultaneously occupying the auxiliary and main operators through DNA looping. The rate of DNA loop formation is $k_{loop} = 660\,\text{molecules}\,k_R^0$ [33]. We also show the effect of DNA looping in the case where the kinetics of dissociation from the looped state are 100 times faster than the kinetics of dissociation from the unlooped state: $c = k_{unloop}/k_R^{off} = 100$. In this limit, the presence of the auxiliary operator leads to less gene expression noise. (D) Prediction for a library of $P_{lacUV5}$ promoter variants, harboring an O2 deletion, and with the position of O3 moved upstream by multiples of 11 bp and keeping its identity or replaced the operator by O$id$. Parameters are taken from the analysis in [33] of the data in [92]. We assume a concentration of 50 lac repressor tetramers per cell. The association rate of the tetrameric repressor to the operators is taken from Table I. The lifetimes of the operator-repressor complex are given in the caption to Figure 2. The dependence of the rate of DNA looping on the interoperator distance is taken from [33], and equal to: $k_{loop} = k_R^{on} \times Exp\left[-\frac{u}{D} - vLog[D] + wD + z\right]$, where $u = 140.6$, $v = 2.52$, $w = 0.0014$, $z = 19.9$.

**Figure 5: Simple activation architecture**. (A) The Fano factor is plotted as a function of the fold-change gene expression (blue line). In red, we show the effect of reducing operator strength (i.e. reducing the lifetime of the operator-activator complex) by a factor of 10. Just as we observed with single repression, weak activator binding operators generate less promoter noise than strong activating operators. The parameters used are shown in Table I with the exception of $r_1 = 0.33\,\text{mRNA s}^{-1}/f$. Inset: Prediction for the activation of the $P_{lac}$ promoter. The fold-change in noise is plotted as a function of the fold-change in mean mRNA expression for both the wild-type $P_{lac}$ (CRP dissociation time =8min), represented by a blue line and a $P_{lac}$ promoter variant where the $lac$ CRP binding site has been replaced by the weaker $gal$ CRP binding site (dissociation time=1 min). The enhancement factor was set to $f = 50$ [33]. These parameters are taken from [68] and [33]. The remaining parameters are taken from Table I. (B) Fano factor as a function of $\langle mRNA \rangle / \langle mRNA \rangle_{max}$ for a repressor and an activator with the same transcription factor affinity. The parameters used are shown in Table I. The transcription rate in the absence of activator is assumed to be zero. The transcription rate in the fully activated case is equal to the transcription rate of the repression construct in the absence of repressor and is $r = 0.33\,\text{mRNA s}^{-1}$ as specified by Table I. For low expression levels $\langle m \rangle / \langle m \rangle_{max} < 0.5$ simple activation is considerably noisier than simple repression. (C) The results of a stochastic simulation for the simple activation and simple repression architectures. We assume identical dissociation rates for the activator and repressor, and identical rates of transcription in their respective active states. As shown in (B), low concentrations of an activator result in few, but very productive transcription events, whereas high concentrations of a repressor lead to the frequent but short lived excursions into the active state.

**Figure 6: Dual activation architecture**. (A) Kinetic mechanism of dual activation. The parameters $k_A^{off}$ and $k_A^{on}$ are the rates of activator dissociation and association to the operators, and $\omega$ is a parameter reflecting the effect of cooperative binding in the dissociation rate. For independent binding, $\omega = 1$ and for cooperative binding $\omega < 1$ (see Table I). (B) Fano factor as a function of the mean mRNA for independent ($\omega = 1$), cooperative ($\omega = 0.1$) and for simple activation. The parameters are taken from Table I and $r_1 = 0.33\,\text{mRNA s}^{-1}/f$, $r_2 = f \times r_1$, $r_3 = f \times r_1$, and $r_4 = f^2 \times r_1$. (C) A stochastic simulation shows the effect of independent and cooperative binding in creating a sustained state of high promoter activity, resulting in high levels of cell-to-cell variability. (D) We present a prediction for the $r1-P_{RM}$ promoter ($P_{RM}$ promoter that does not exhibit $O_{R3}$ mediated repression [51]). This promoter is activated by cI, which binds cooperatively to $O_{R1}$ and $O_{R2}$. The prediction is shown for wild-type cI ($\omega = 0.013$) and for a cooperativity deficient mutant (Y210H, $\omega = 1$). Parameters taken from [33,43,59,91]. The lifetime of $O_{R1}$-cI complex is 4min. Lifetime of $O_{R2}$-cI complex is 9.5s.

# TABLES AND CAPTIONS

| Kinetic Rate | Symbol | Value | Reference |
|---|---|---|---|
| Unregulated promoter transcription rate | $r$ | $r = 0.33 \, \text{mRNA} \, \text{s}^{-1}$ | [90] |
| Repressor and activator associations rates | $k_R^0, k_A^0$ | $0.0027 \, (\text{s molecule})^{-1}$ | [7] |
| Repressor and activator dissociation rates | $k_R^{off}, k_A^{off}$ | $0.0023 \, \text{s}^{-1}$ | [42] |
| mRNA decay rate | $\gamma$ | $0.011 \, \text{s}^{-1}$ | [15] |
| Ratio between transcription rates due to activation | $f = r_1/r_2$ | 11 | [51] |
| Cooperativity in repression | $\omega_{repression}$ | 0.013 | [51] |
| Cooperativity in activation | $\omega_{activation}$ | 0.1 | [33] |
| Looping J-factor | $[J]$ | 660 molecules | [33] |
| Protein translation burst size | $b$ | 31.2 proteins/mRNA | [10] |
| Protein decay rate | $\gamma_{protein}$ | $0.00083 \, \text{s}^{-1}$ | [90] |

**Table I: Kinetic parameters used to make the quantitative estimates in the text and plots in the figures.** These parameters are all measured for model systems such as the $P_{lac}$ promoter or the $P_{RM}$ in *E. coli*, and are here considered representative for promoter-transcription factor interactions. Specific promoter architectures of interest to readers will demand careful attention to the determination of particular parameter values for the promoter of interest.

| Promoter architecture | Fold-change in noise |
|---|---|
| 1. Simple repression 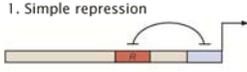 | $1 + \dfrac{r\,k_R^{on}}{(k_R^{off} + k_R^{on})(\gamma + k_R^{off} + k_R^{on})}$ |
| 2. Simple activation 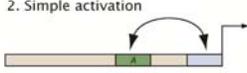 | $1 + \left( \dfrac{\left(\frac{r_2}{r_1} - 1\right)^2 k_A^{off} k_A^{on} r_2}{(k_A^{off} + k_A^{on})(\gamma + k_A^{off} + k_A^{on})\left(k_A^{off} + \frac{r_2}{r_1} k_A^{on}\right)} \right)$ |
| 3. Dual repression 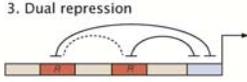 | $1 + \dfrac{\left(r\,k_R^{on}\left(k_R^{2\,on} + 2\omega k_R^{off}(\gamma + 2w\,k_R^{off}) + k_R^{on}(\gamma + k_R^{off} + 4\omega k_R^{off})\right)\right)}{\left(\left(2(k_R^{on})^2 + (\gamma + k_R^{off})(\gamma + 2\omega k_R^{off}) + k_R^{on}(3\gamma + 4\omega k_R^{off})\right)\left((k_R^{on})^2 + \omega k_R^{off}(k_R^{off} + 2k_R^{on})\right)\right)}$ |
| 4. Cooperative activation 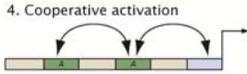 | $1 + \dfrac{(r_2/r_1 - 1)^2 r_2\,\omega\,k_A^{off} k_A^{on}}{\omega k_A^{off}(k_A^{off} + k_A^{on}) + r_2/r_1\,k_A^{on}(\omega k_A^{off} + k_A^{on})} \left( \dfrac{1}{2(\gamma + k_A^{off} + k_A^{on})} \right.$ $\left. + \dfrac{2(k_A^{on})^3 + (k_A^{on})^2(\gamma + 6k_A^{off}) + \omega(k_A^{off})^2(\gamma + 2\omega k_A^{off}) + 2k_A^{off} k_A^{on}(\gamma + k_A^{off} + 2\omega k_A^{off})}{2\left(2(k_A^{on})^2 + (\gamma + k_A^{off})(\gamma + 2\omega k_A^{off}) + k_A^{on}(3\gamma + 4\omega k_A^{off})\right)\left((k_A^{on})^2 + \omega k_A^{off}(k_A^{off} + 2k_A^{on})\right)} \right)$ |
| 5. Repression by DNA looping 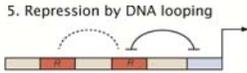 | $1 + r\,k_R^{on} k_l \dfrac{(k_R^{off} + k_R^{on})(\gamma + k_R^{off} + k_R^{on})\left(\gamma + 2(k_R^{off} + k_R^{on})\right)}{(\gamma + k_l + k_R^{off} + k_R^{on})\left(k_l k_R^{on} + (k_R^{off} + k_R^{on})^2\right)}$ $\dfrac{1 + \left(k_l k_R^{off} + 2(k_R^{on})^2 + 2k_R^{off}(\gamma + 2k_R^{off}) + k_R^{on}(\gamma + 5k_R^{off})\right)\left((k_R^{off} + k_R^{on}) + \left(\gamma^2 + 4(k_R^{off} + k_R^{on})^2 + \gamma(5k_R^{off} + 4k_R^{on})\right)\right)}{k_l(\gamma + 2k_R^{on}) + (\gamma + k_R^{off} + k_R^{on})\left(\gamma + 2(k_R^{off} + k_R^{on})\right)}$ |

**Table II: Fold-change in noise for different promoter architectures.** The fold-change in promoter noise is shown as a function of the different kinetic parameters corresponding to each promoter architecture considered throughout the text. Refer to Table I for the definition and value of each rate.

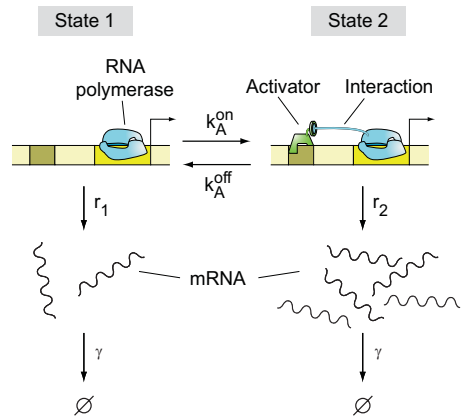
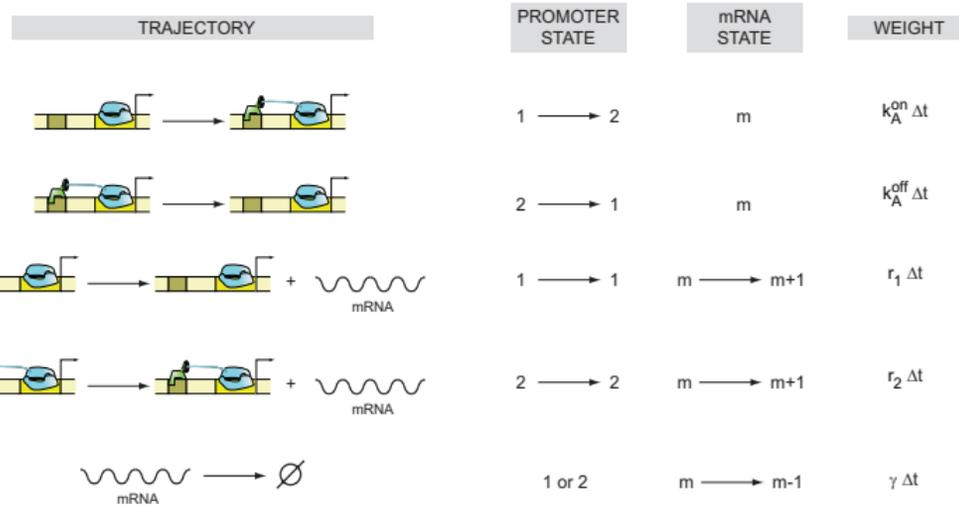
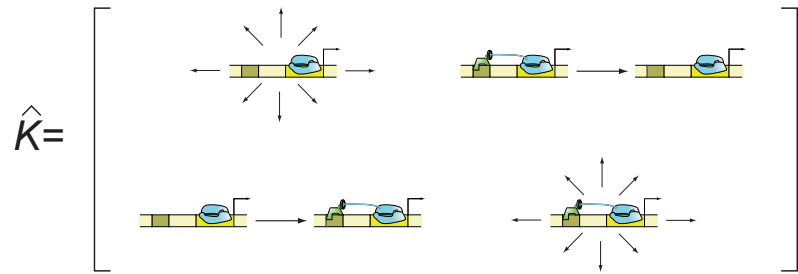
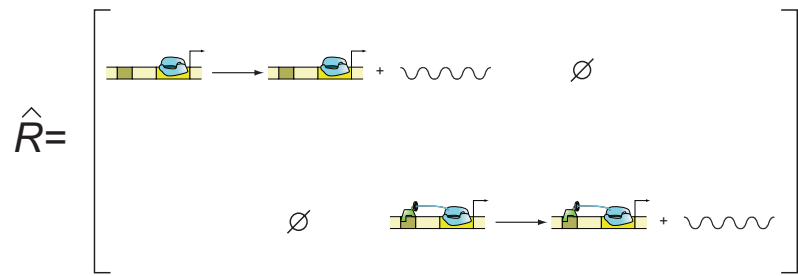
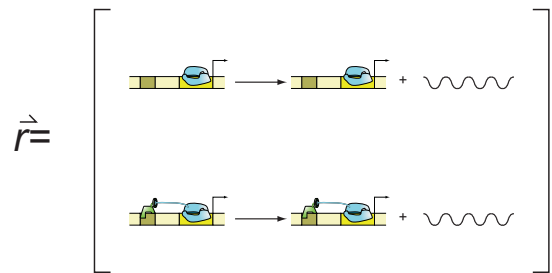

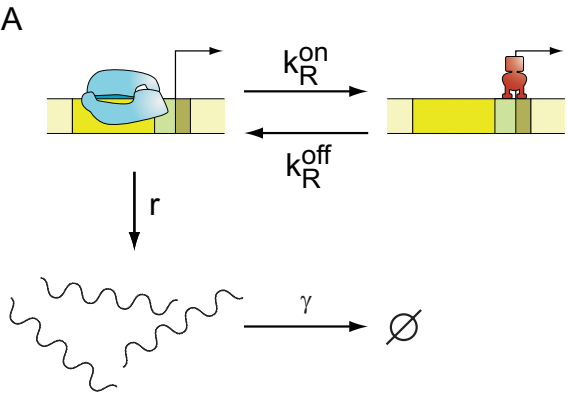
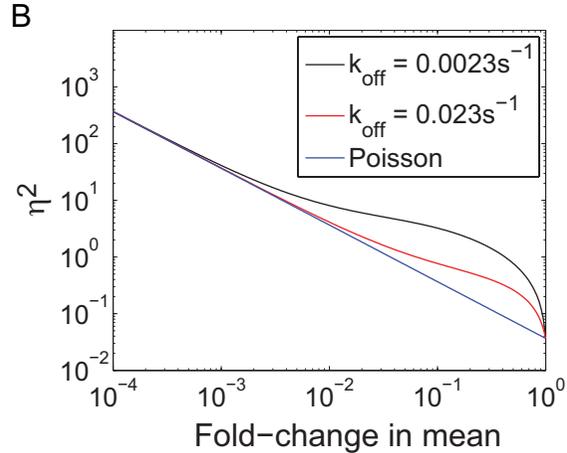
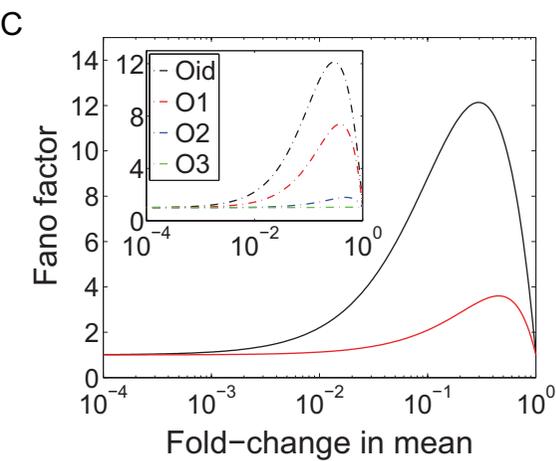
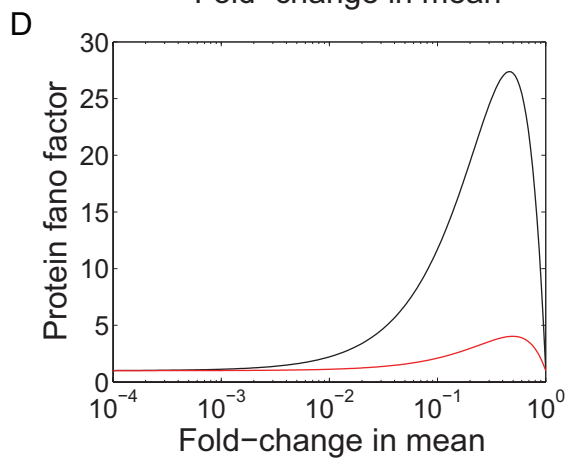
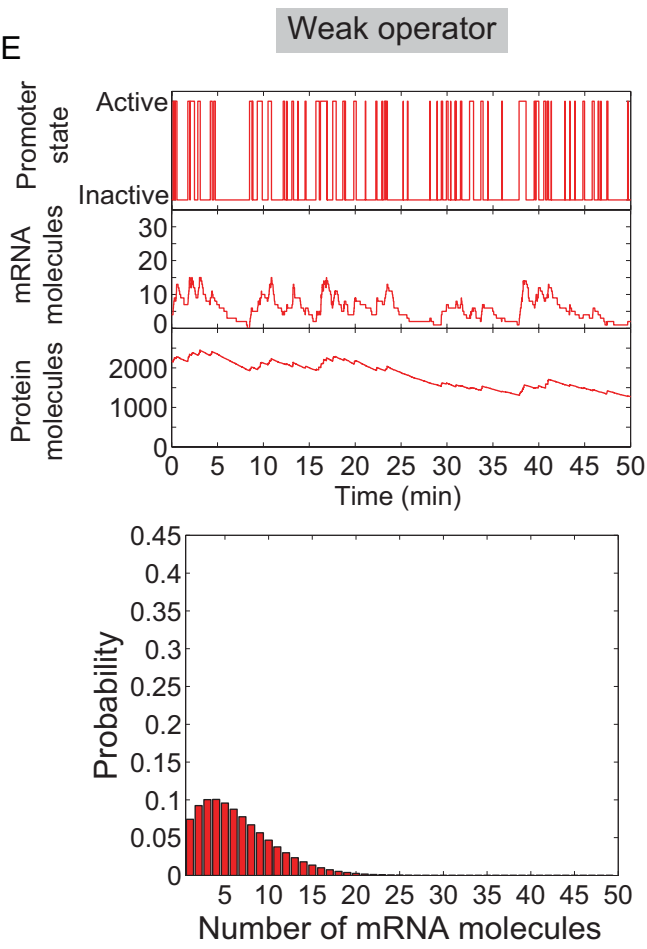
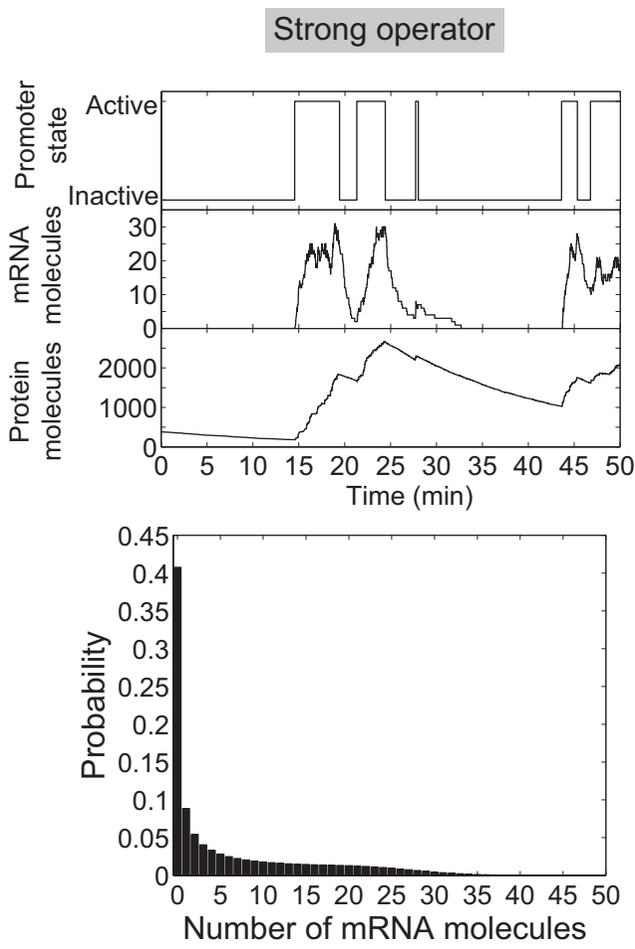

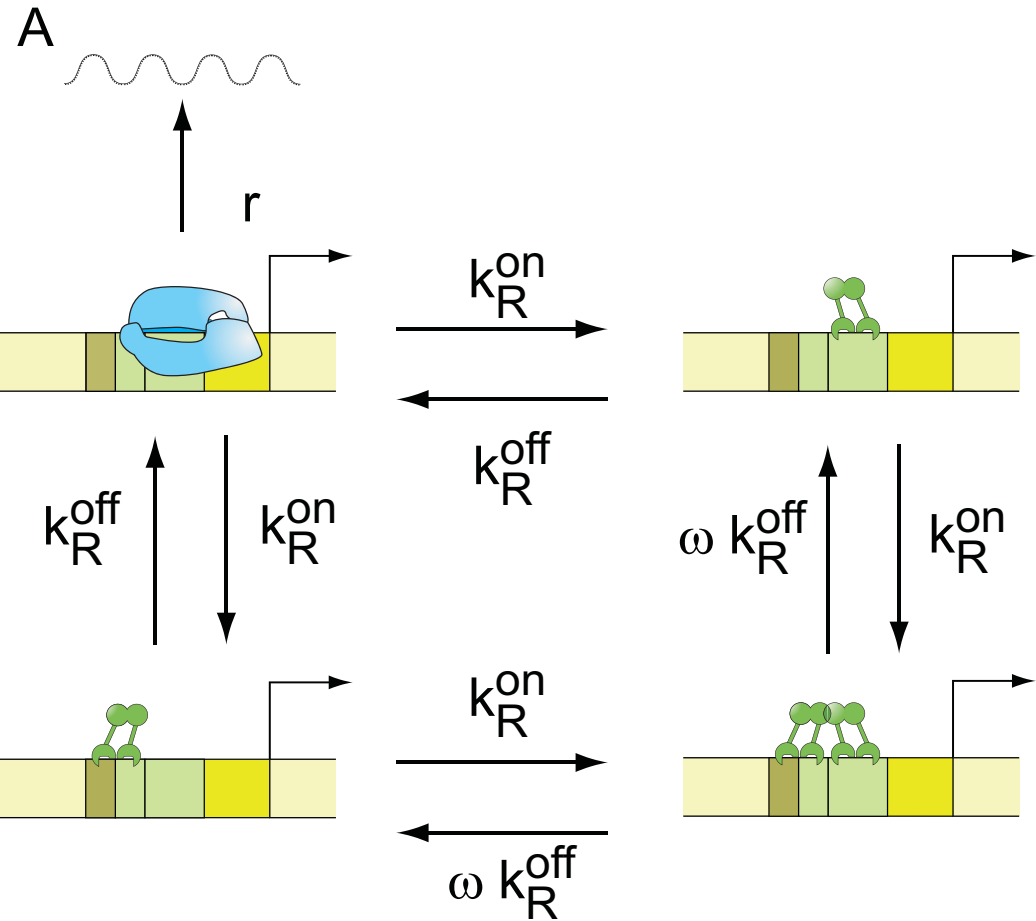

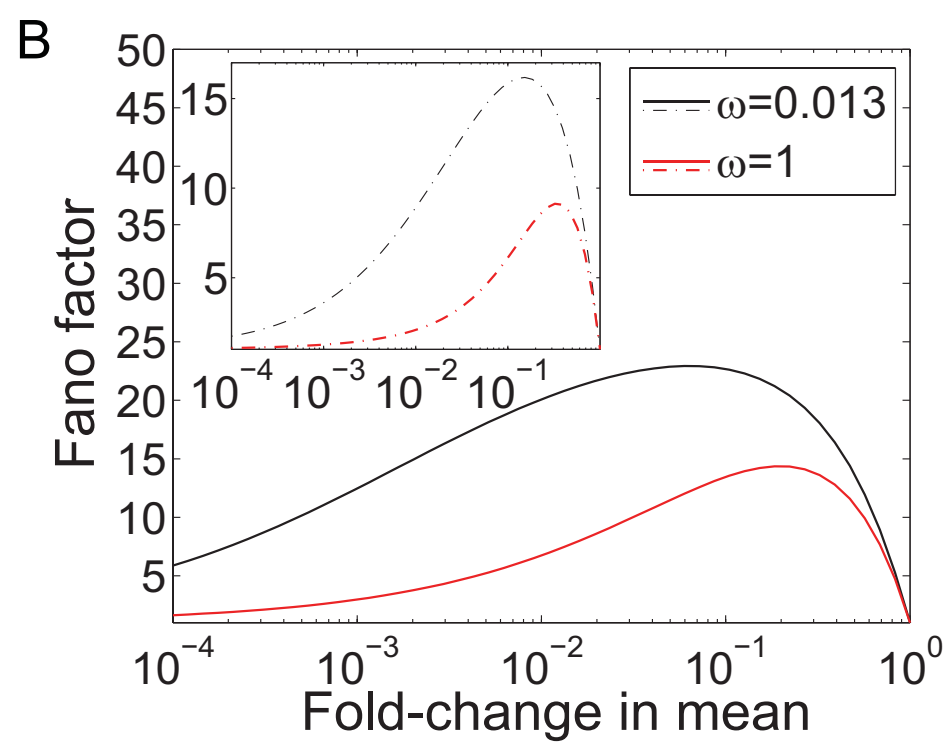

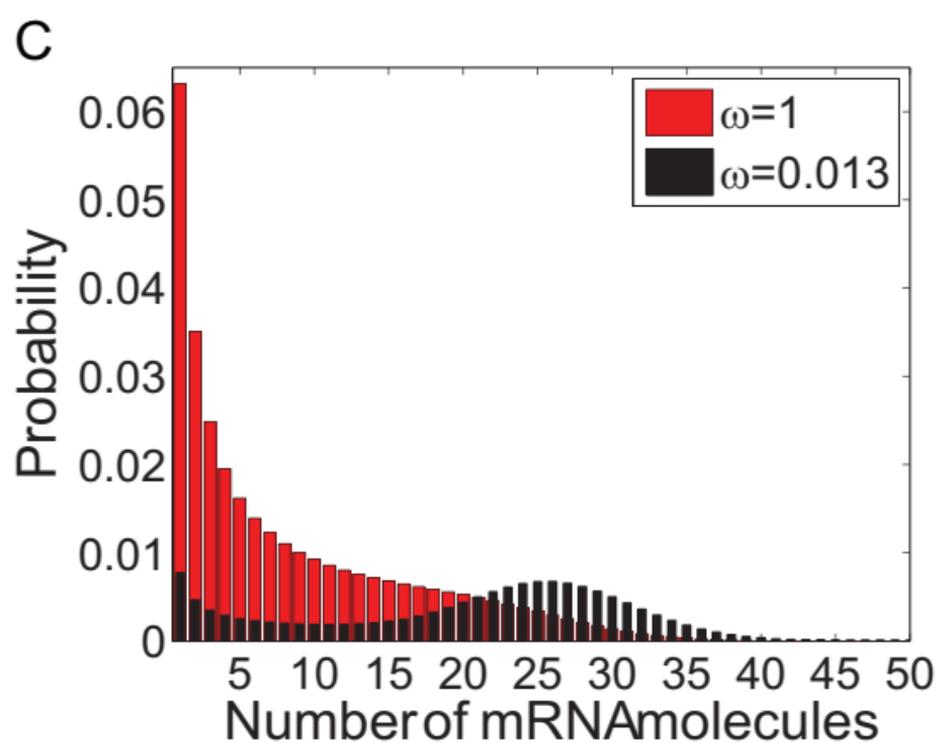

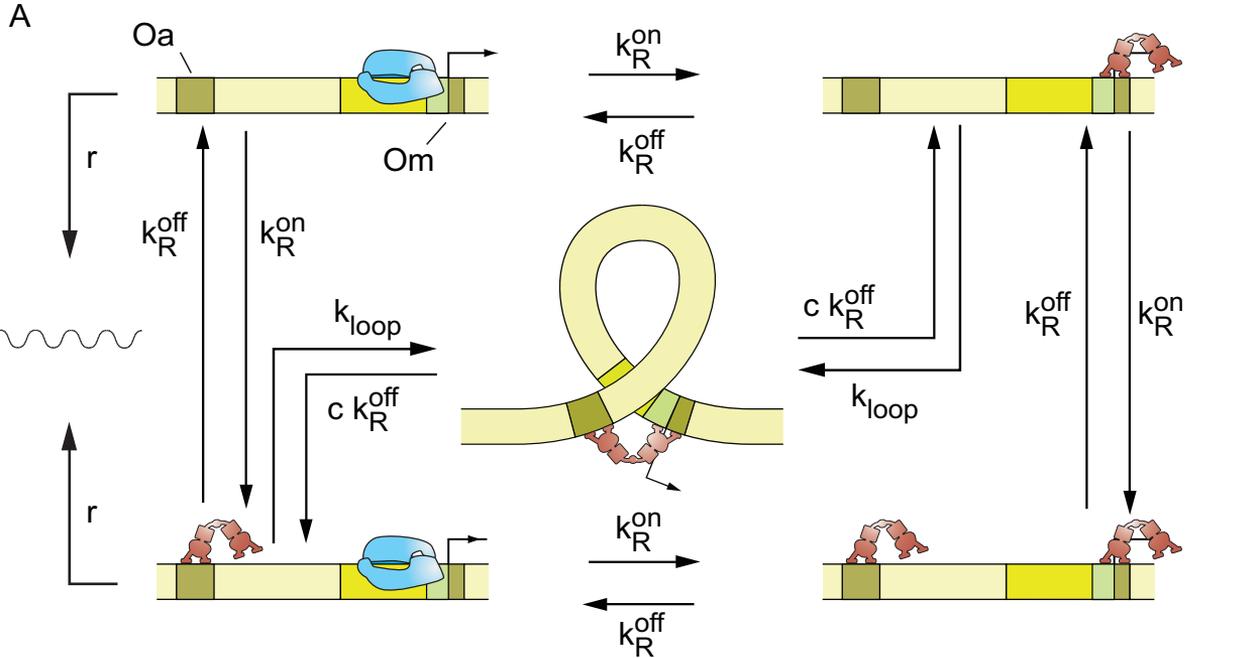

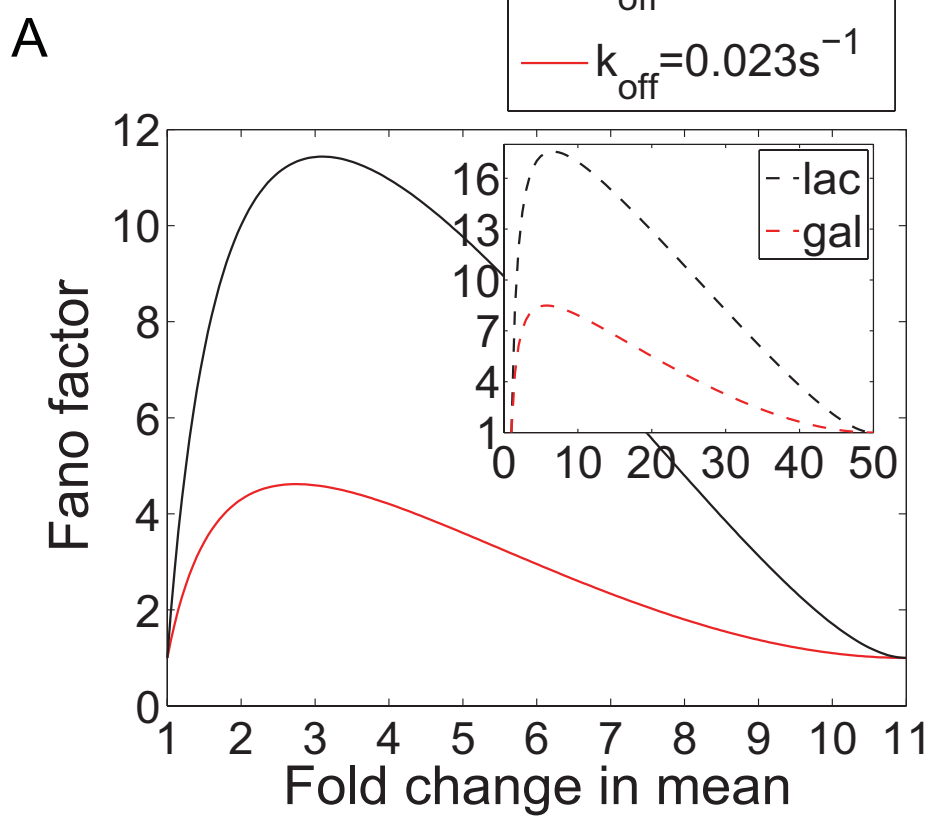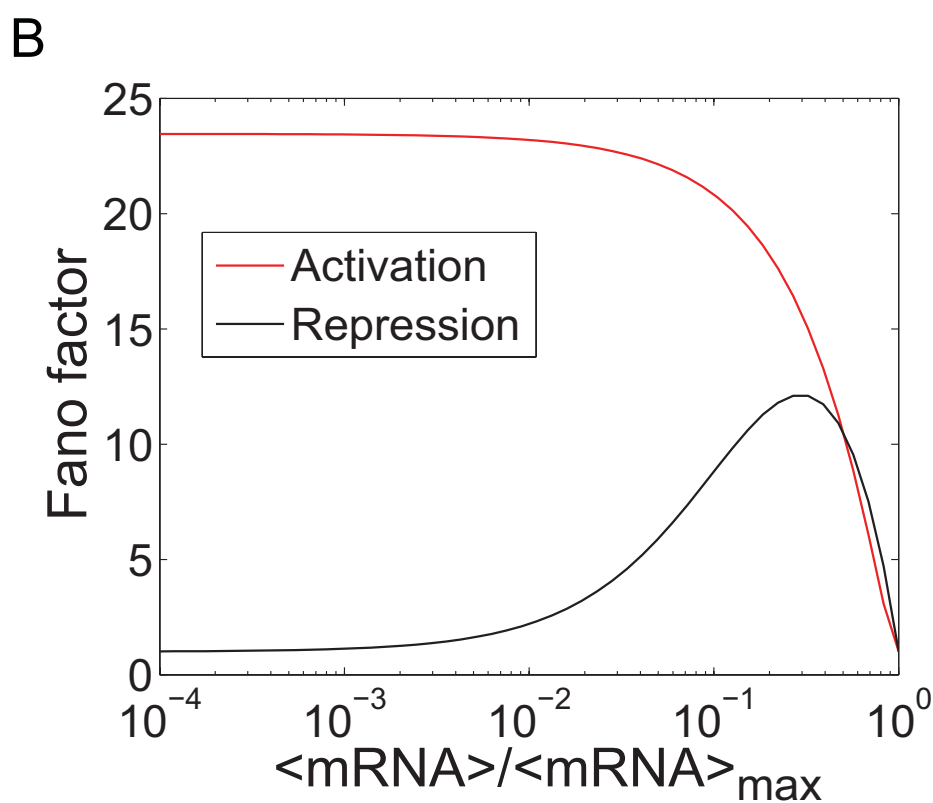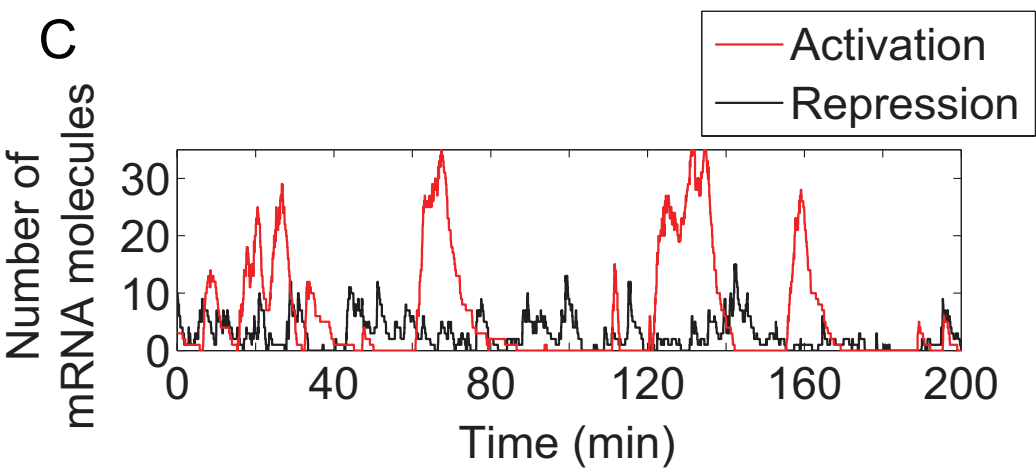

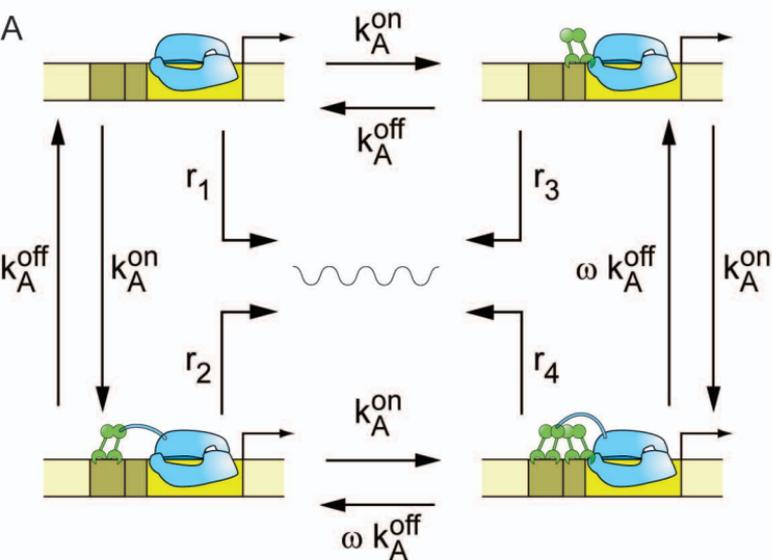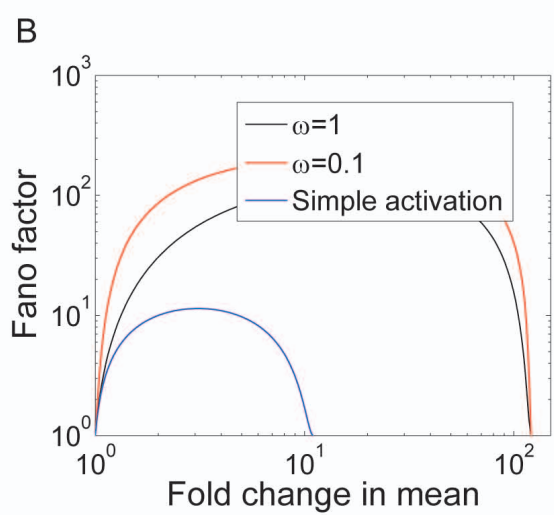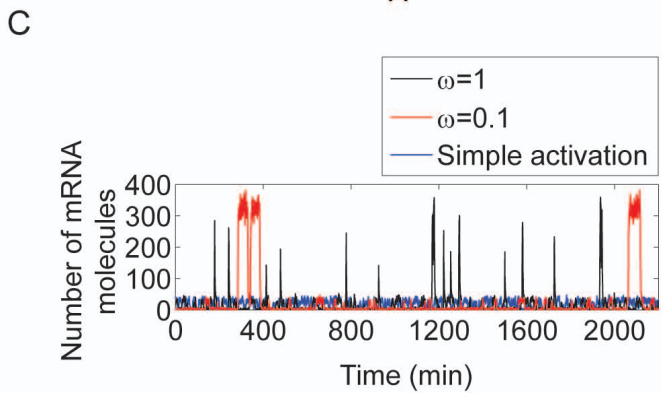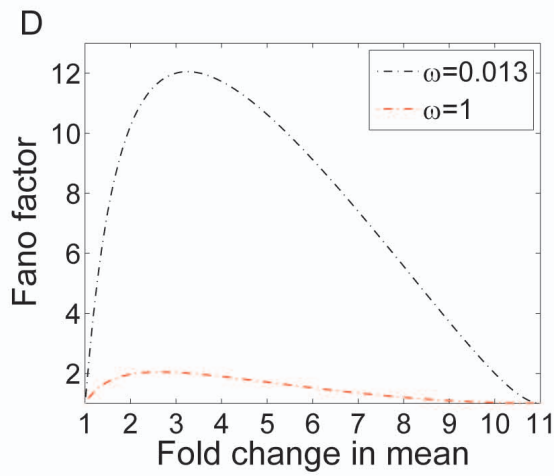